\documentclass[sigplan,10pt,screen]{acmart}
\renewcommand\footnotetextcopyrightpermission[1]{}
\usepackage{xspace}
\usepackage{tabularx}
\usepackage{subfig}
\usepackage{enumitem}
\usepackage{algorithm}
\usepackage{algorithmicx}
\usepackage[noend]{algpseudocode}
\usepackage{enumitem}
\usepackage{pifont}
\setitemize{noitemsep,topsep=0pt,parsep=0pt,partopsep=0pt}

\newcommand{\parabf}[1]{\medskip\noindent\textbf{#1}}

\newcommand{\cut}[1]{}

\newcommand{\sysname}{LayerKV\xspace}

\begin{document}
\title{\sysname: Optimizing Large Language Model Serving with Layer-wise KV Cache Management}
\pagestyle{plain}

\author{
\vspace{-0.1in}
Yi Xiong$^{\ast,\ddagger}$\qquad Hao Wu$^{\ast}$\qquad Changxu Shao$^{\ast}$\qquad
Ziqing Wang\qquad Rui Zhang\qquad Yuhong Guo\qquad Junping Zhao$^{\dagger}$\qquad
Ke Zhang\qquad Zhenxuan Pan\\
\textit{Ant Group} 
}

\thanks{$\ast$ Co-first authors. Yi Xiong, Hao Wu, Changxu Shao: \{alex.xy, wh391609, shaochangxu.scx\}@antgroup.com. }
\thanks{$\dagger$ Corresponding author. Junping Zhao: junping.zjp@antgroup.com. }
\thanks{$\ddagger$ Work done during an internship at Ant Group}

\begin{abstract}
The expanding context windows in large language models (LLMs) have greatly enhanced their capabilities in various applications, but they also introduce significant challenges in maintaining low latency, particularly in Time to First Token (TTFT). 
This paper identifies that the sharp rise in TTFT as context length increases is predominantly driven by queuing delays, which are caused by the growing demands for GPU Key-Value (KV) cache allocation clashing with the limited availability of KV cache blocks.
To address this issue, we propose \sysname, a simple yet effective plug-in method that effectively reduces TTFT without requiring additional hardware or compromising output performance, while seamlessly integrating with existing parallelism strategies and scheduling techniques. 
Specifically, \sysname introduces layer-wise KV block allocation, management, and offloading for fine-grained control over system memory, coupled with an SLO-aware scheduler to optimize overall Service Level Objectives (SLOs).
Comprehensive evaluations on representative models, ranging from 7B to 70B parameters, across various GPU configurations, demonstrate that \sysname improves TTFT latency up to 69x and reduces SLO violation rates by 28.7\%, significantly enhancing the user experience.
\end{abstract}

\maketitle


\section{Introduction}
\label{sec:introduction}

The advent of large language models (LLMs) has ushered modern applications into a new era, characterized by significant advancements across various domains such as coding assistants~\cite{nijkamp2023codegen}, conversation~\cite{chatgpt}, and planning ~\cite{significantgravitasautoGPT}.
A critical feature of LLMs is their context window, which is rapidly expanding to enable advanced analysis of extensive documents, effective problem-solving within large codebases, and customized content generation based on detailed instructions~\cite{gemini}. 
Notable examples include Anthropic’s Claude-3~\cite{claude3}, Google’s Gemini-1.5~\cite{gemini}, and UC Berkeley’s Large World Model (LWM)~\cite{liu2023world}, all of which support a context window of up to 1 million tokens.

\begin{figure}[t]
    \centering
    \subfloat[Average TTFT and TPOT Across Varying Context Lengths.\label{fig:ttft-a}]{
        \includegraphics[width=0.45\textwidth]{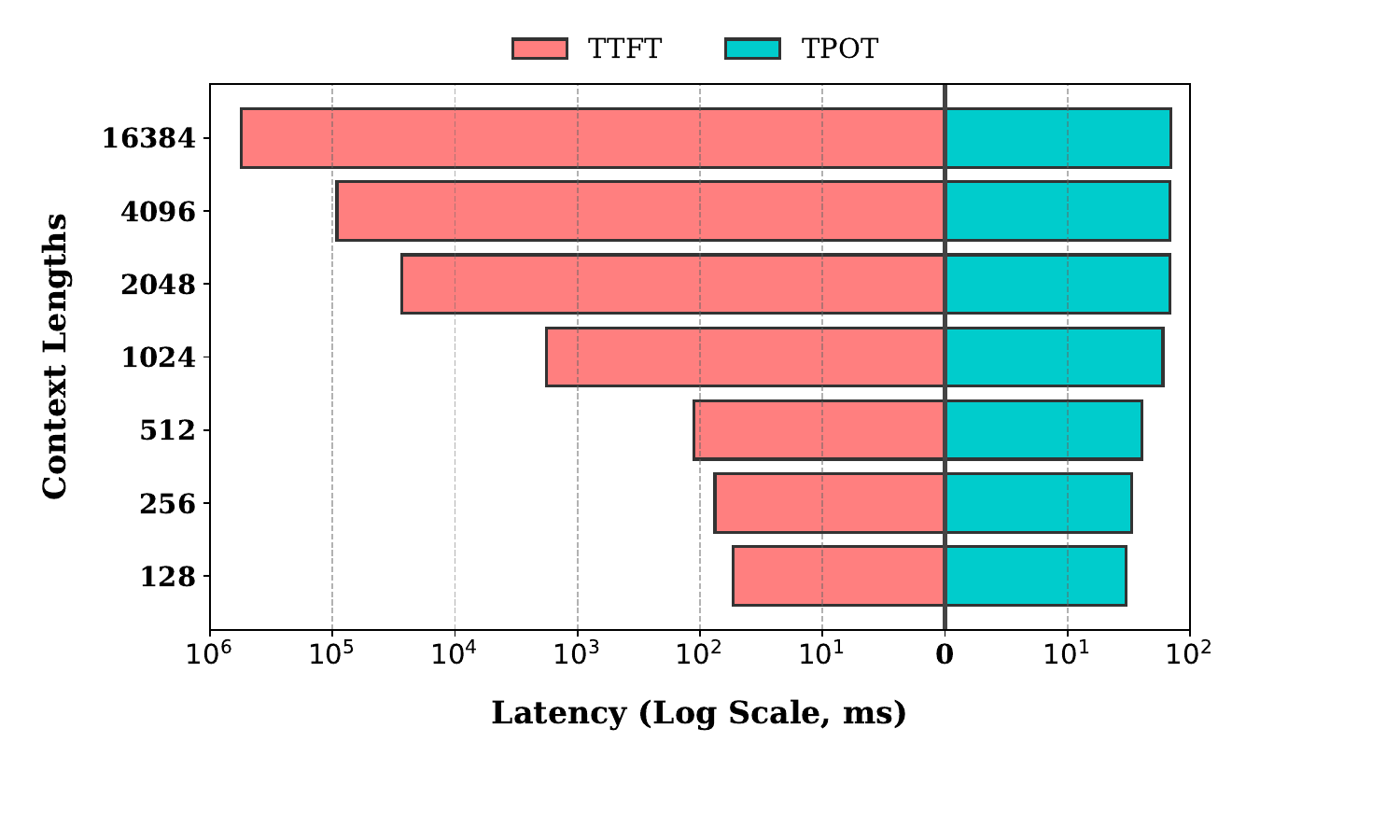}
    }\\
    \subfloat[Breakdown of Queuing and Prefill Latencies within TTFT.\label{fig:ttft-b}]{
        \includegraphics[width=0.45\textwidth]{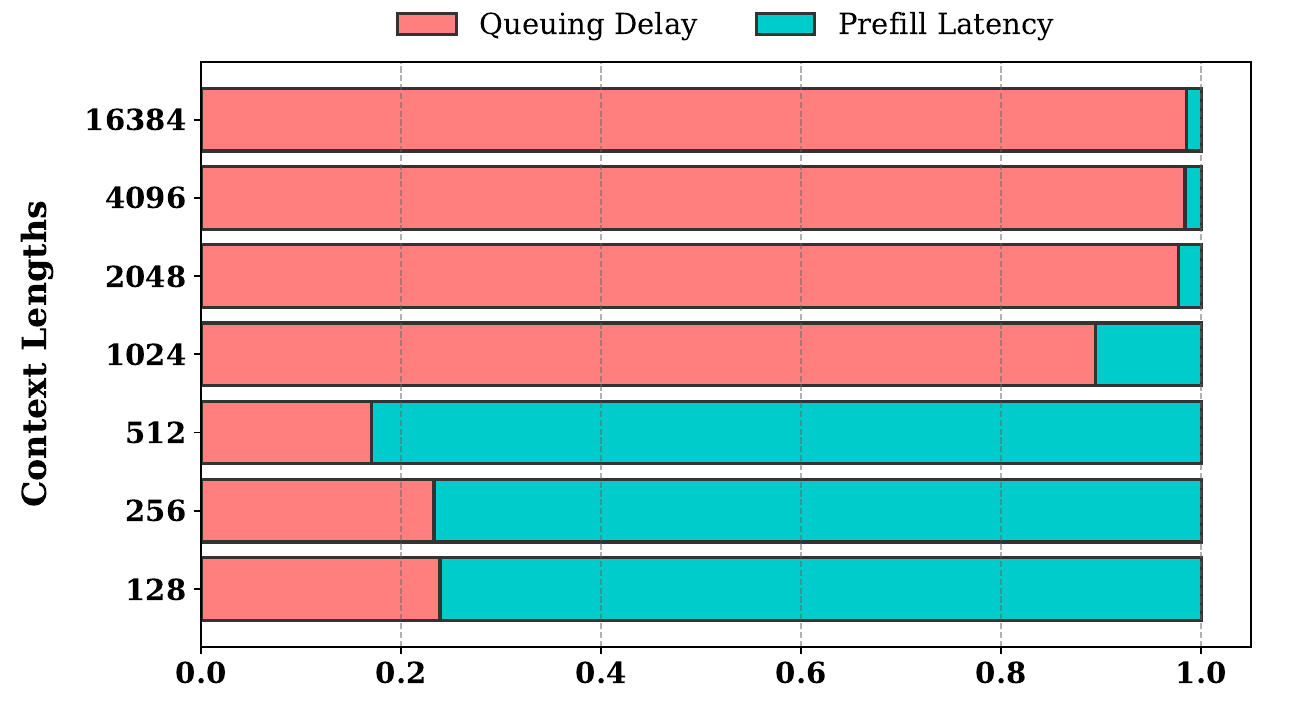}
    }
    \caption{LLaMA-2-7B~\cite{DBLP:llama2} on a single L20 GPU with 48GB memory at a request arrival rate of 1 req/s. All latency measurements represent the average across 100 requests.}
    \label{fig:ttft}
\end{figure}

The increasing context length introduces challenges in maintaining smooth live interactions, as the user experience in LLM serving is directly impacted by token generation latency.
Specifically, various metrics can be used to measure Service Level Objectives (SLOs).
The most critical SLO metrics are Time to First Token (TTFT), which measures the latency from request arrival to the generation of the first token, encompassing both queuing delay\footnote{Waiting for prefill schedule.} and prefill latency; and Time Per Output Token
(TPOT\footnote{Also known as Time Between Tokens (TBT) or Inter-Token Latency (ITL).}), defined as the average time between consecutive tokens for the same request.
However, as context length increases, TTFT becomes dramatically prone to violating SLO requirements.
This is clearly demonstrated by an experiment where the prompt length was increased from 128 to 16k tokens while keeping the output length fixed at 512 tokens.
As figure~\ref{fig:ttft} shows: (1) TTFT exhibits a quadratic increase as context length extends, while TPOT scales linearly. Thus, reducing average TTFT is pivotal in addressing long-context challenges. (2) As the context length increases to 1024, queuing delay becomes the dominant factor in TTFT. This insight directs our optimization efforts towards reducing queuing delay.



Some existing approaches can be employed to further reduce the TTFT, which can be broadly categorized into three types:
\textbf{Parallel Computation}: These methods distribute computation across multiple cores or devices, significantly accelerating inference.
Notable examples include sequence parallelism~\cite{DBLP:ring-attention,DBLP:striped-attention}, tensor parallelism~\cite{DBLP:Megatron-LM,DBLP:conf/mlsys/PopeDCDBHXAD23}, pipeline parallelism~\cite{DBLP:mooncake,DBLP:alpa,agrawal2023sarathi} and prefill-decoding disaggregation~\cite{zhong2024distserve}. 
Nonetheless, these approaches often necessitate additional hardware investment.
\textbf{Algorithmic Innovations}: This category includes model lightweighting techniques such as sparsity~\cite{DBLP:dejavu,DBLP:MInference} and quantization~\cite{DBLP:gptq,DBLP:awq}, aimed at reducing the memory footprint and computational demands of LLMs by creating more efficient and compact models. 
Additionally, attention variants like window attention~\cite{DBLP:streaming-llm,DBLP:h2o}, linear attention~\cite{DBLP:Linear-Attention,DBLP:journals/corr/abs-2101-10277}, and activation-shared attention~\cite{DBLP:GQA,DBLP:MQA} introduce novel architectures beyond the traditional Transformer, enabling faster and more resource-efficient inference. 
However, these methods often compromise the model’s output quality.
\textbf{Request Scheduling}: These methods aim to reduce queuing delays by employing preemption~\cite{DBLP:conf/osdi/HanZ0022} or by prioritizing jobs based on their expected completion times~\cite{DBLP:journals/corr/abs-2305-05920, DBLP:mooncake, DBLP:journals/corr/abs-2404-08509}. 
However, these methods face challenges in maintaining fairness among requests, which can even lead to starvation.

Unlike prior studies, this paper investigates the underlying causes of the sharp increase in queuing time as context length varies.
Specifically, we observe that this phenomenon stems from the growing allocation demands for KV cache, which clash with limited GPU KV cache blocks, as detailed in (§~\ref{sec:background}).
To this end, we propose \sysname, a simple yet effective layer-wise KV cache offloading method, that significantly reduces TTFT without introducing additional hardware, sacrificing performance, or causing starvation, while still ensuring TPOT requirements. 
Moreover, \sysname is fully compatible with the aforementioned methods, enabling further TTFT optimization if needed.


In summary, we make the following contributions:
\begin{itemize}[leftmargin=*]
    \item  We identify that queuing delays significantly affect TTFT SLO due to the conflict between the growing KV cache memory demand and the limited GPU KV block resources, ultimately degrading the user experience.
    \item We design \sysname, which introduces layer-wise KV block allocation, management, and offloading for fine-grained control over system memory, coupled with an SLO-aware scheduler to optimize overall SLOs.
    \item We conduct comprehensive evaluations on models ranging from 7B to 70B across single and multiple GPUs, demonstrating significant optimizations. \sysname achieves up to 69x improvements in TTFT, reducing Service SLO violation rates by 28.7\%, thereby significantly enhancing the user experience.
\end{itemize}


\section{Background and Motivations}
\label{sec:background}
\begin{figure*}[h]
     \includegraphics[width=\textwidth]{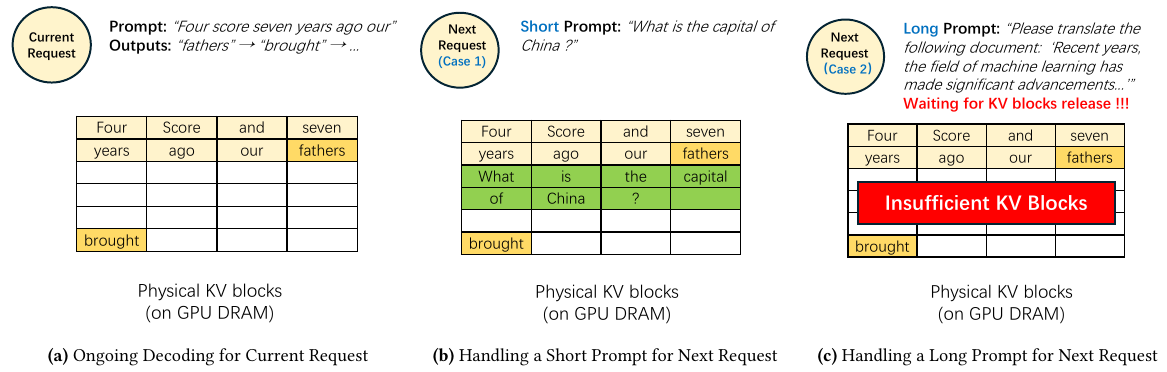}
    \caption{The surge in queuing delays is caused by the inability to process long prompts due to insufficient KV blocks.}
    \label{fig:three_subfigs}
\end{figure*}


\subsection{Preliminary}


\subsubsection{The Process of LLM Inference}

Most of the popular LLMs~\cite{openai2023gpt4,DBLP:llama2} are built upon the decoder-only transformer architecture~\cite{vaswani2017attention}. 
These models consist of stacked transformer layers, each containing an attention mechanism and a feed-forward network (FFN). 
The attention layers facilitate token interactions within a request, while the FFN processes tokens individually. 
During each iteration, given the preceding tokens, the model predicts the next token.

To prevent redundant computations, LLM inference stores the keys and values of all attention layers from preceding tokens in GPU memory, referred to as the KV cache, which can be frequently reused for subsequent token generation. 
This optimization splits the generation process into two phases: the \textit{prefill phase} and the \textit{decoding phase}.

In the prefill phase, all input tokens are processed in parallel to generate the initial output token. 
The ability to process input tokens concurrently in this phase typically results in high computational demands, except for requests with short prompts. 
Since the computational complexity of attention mechanisms scales quadratically with input length, while that of FFNs scales linearly, the computation time in the prefill phase generally grows superlinearly with input length. 
In contrast, the decoding phase only produces the key-value cache for the newly generated output token.

\subsubsection{Existing LLM Serving Systems}
The compute utilization in serving LLMs can be improved by batching multiple requests. 
Because the requests share the same model weights, the overhead of moving weights is amortized across the requests in a batch. 
For LLMs that have variable-sized input and output, the granularity of batching has a huge impact on system throughput and serving latency. 
If scheduling is performed at the request granularity, executing a batch of requests with different input prompt lengths requires padding tensors to the maximum length and waiting for the request with the longest output to finish. Iteration-level batching strategy, originally proposed by BatchMaker~\cite{DBLP:batchmaker} for non-transformer-based sequence-to-sequence models, performs batching at token granularity. 
ORCA~\cite{DBLP:orca} extends this approach to support the LLM workload: whenever a request finishes an iterative decoding step, the scheduler checks whether it has reached the end of a sequence and can leave the batch, making room for requests to start their computation immediately.

For each request, the model performs iterative generation until either the special end-of sentence token (EOS) is emitted or the preconfigured maximum decoding length is reached. 
However, LLM serving systems like ORCA~\cite{DBLP:orca} and FasterTransformer~\cite{nvidia2019fastertransformer} pre-allocate slots in the KV cache for each request based on the maximum possible decoding length, leading to inefficient memory usage. 
In contrast, PagedAttention~\cite{kwon2023efficient} dynamically adjusts the size of cache slots for each request as needed and allows these slots to be stored in non-contiguous GPU memory. 
Advanced LLM serving systems, such as vLLM~\cite{kwon2023efficient}, integrate the aforementioned techniques for request scheduling and KV cache management, respectively.

\subsection{Motivation}
As depicted in Figure~\ref{fig:ttft}, due to its superlinear increase in latency, TTFT increasingly struggles to meet SLO requirements as the context length grows. 
Notably, this surge is predominantly driven by queuing delay, rather than the more widely discussed prefill latency.
We conducted an in-depth analysis of the reasons behind the sharp increase in average queuing delays as the context length extends and visualized in Figure~\ref{fig:three_subfigs}.

Firstly, Figure~\ref{fig:three_subfigs} (a) illustrates the current phase, where the system leverages PagedAttention GPU kernel to handle decoding iterations with non-contiguous stored KV caches.
PagedAttention reserves a significant portion of GPU memory for KV blocks, intended to store future KV cache entries.
To determine the amount of memory to allocate for KV blocks, the system profiles the available GPU memory during initialization based on the maximum configured input size. During this process, a fixed proportion (e.g., 90\%) of the remaining memory—after accounting for model parameters and activations—is reserved for KV blocks.
As context window becomes longer, maximum input configurations correspondingly expands, resulting in greater activation memory usage during profiling. 
Consequently, the GPU memory for KV blocks decreases.
This figure deliberately displays a small number of KV blocks to highlight that the capacity of KV blocks can be significantly limited.

Secondly, Figures~\ref{fig:three_subfigs} (b) and (c) illustrate scenarios where requests of varying lengths are queued for scheduling.
Existing serving systems are stateless across requests.
In other words, they de-allocate all the cache slots used by a request as soon as it finishes.
Therefore, within the iterative batch processing approach, the system allows new requests to initiate the prefill stage earlier if sufficient KV blocks are available, thus reducing queuing delays.
To determine whether a new request can be inserted, the system compares the total KV blocks required for its prefill stage with the currently available KV blocks. 
As a result, if requests have shorter prompts, they can be scheduled immediately, as indicated by the green segment. 
However, if requests have longer prompts, they must wait until KV blocks are released, necessitating at least one request to be fully completed first.
This process can be time-consuming, which consequently results in subsequent requests remaining queued.

In summary, the significant rise in average queuing delays is caused by the increasing allocation requirements for KV cache, which come into conflict with the limited number of GPU KV cache blocks.

\section{Desgin}
\label{sec:design:mechanism}

\begin{figure}[h]
    \centering
    \includegraphics[width=0.5\textwidth]{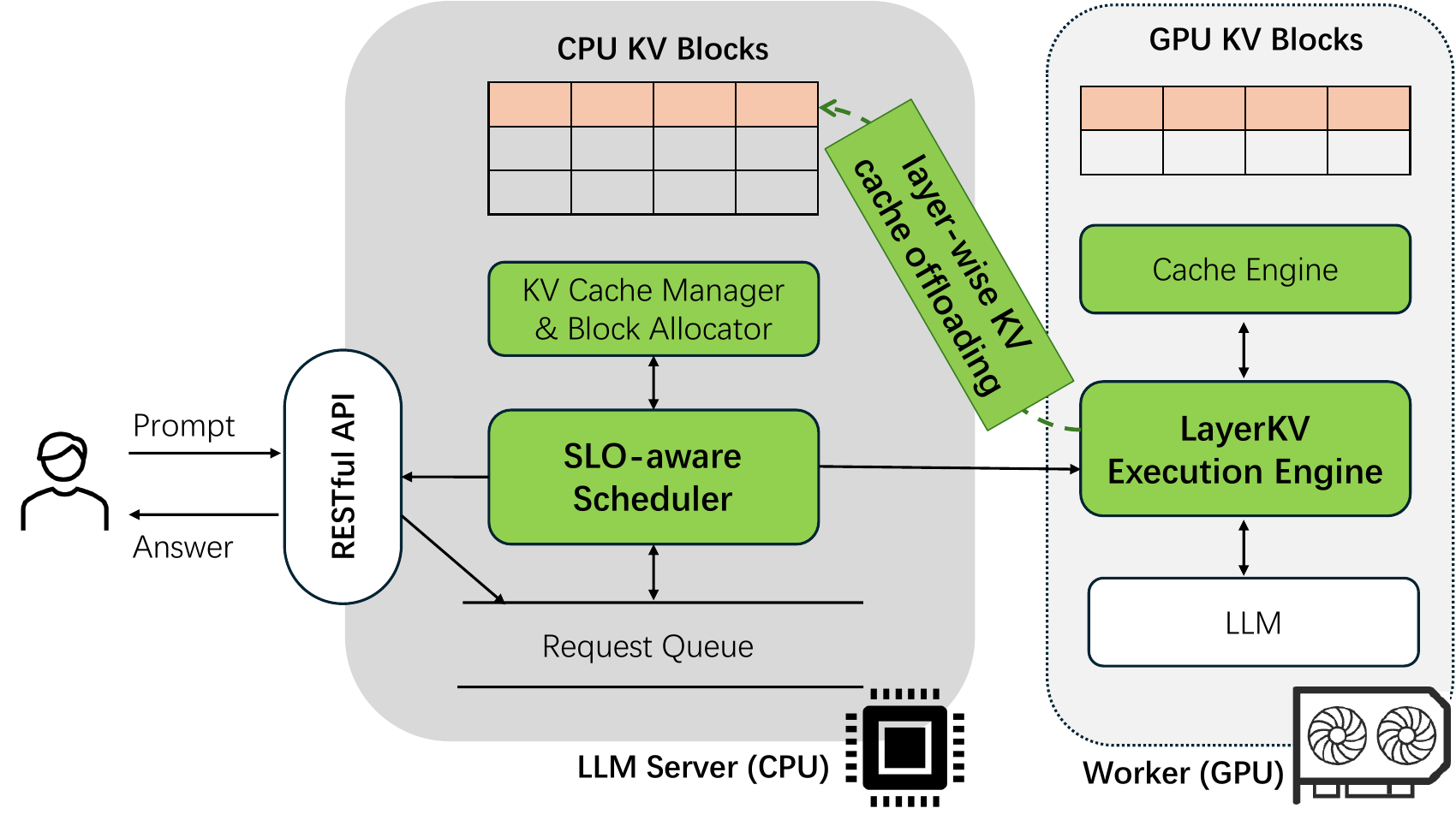} 
    \caption{LayerKV System Overview}
    \label{fig:framework} 
\end{figure}

To mitigate queuing delays caused by limited GPU KV block resources when handling long prompts, our core idea is to refine the granularity of the KV cache to a layer-wise level, rather than retaining the entire KV cache of the prompt in GPU KV blocks.
By implementing layer-wise KV block allocation and KV cache offloading, the demand for GPU KV blocks is reduced, which in turn facilitates the scheduling of new requests. 
This reduction in queuing delays directly contributes to the optimization of the TTFT SLO.

However, concrete design of this idea is nontrivial, as an imprudent approach could result in TPOT SLO violations or a decrease in the number of queries per second (QPS).
Specifically, inserting prefill stage during the current decoding stages can optimize the TTFT of queued requests, but this may lead to an increase in TPOT for requests currently being decoded.
Therefore, determining how to optimize TTFT while satisfying the TPOT SLO is a key design consideration.
Furthermore, the system’s QPS is closely tied to the efficient utilization of computing resources.  
If computation is stalled due to additional PCIe communication, QPS may decrease, negatively impacting overall system throughput. 
Therefore, managing PCIe communication effectively without compromising computational efficiency is another crucial design consideration.

Based on the above analysis, we design LayerKV, a simple and lightweight plug-in for existing LLM inference service systems that effectively optimizes TTFT while ensuring compliance with TPOT SLO and maintaining QPS of systems.
The overall architecture of LayerKV is illustrated in Fig.~\ref{fig:framework}.
The SLO-aware Scheduler directly addresses the first key design consideration, determining whether and how many requests’ prefill stages can be scheduled earlier, ensuring the optimization of average TTFT without compromising the TPOT SLO of requests that still need to generate tokens.
Moreover, the LayerKV Execution Engine and layer-wise KV cache offloading are key components that handle computation and communication processes, while the KV Cache Manager, Block Allocator, and Cache Engine manage the logical and physical aspects of the layer-wise KV cache.
These components address the second design consideration, ensuring that LayerKV’s design optimizes TTFT with almost no negative impact on QPS.

\subsection{SLO-aware Scheduler}
\label{subsec:slo}
The primary task of the SLO-aware Scheduler is to determine the maximum number of prefill phases that can be scheduled without violating the TPOT SLO of requests currently in the decoding phase. 
This is achieved by analyzing both the historical and future states of all decoding requests.
For any given request in the decoding phase, the decision-making process considers the historical decoding time and the number of tokens already decoded, as well as the projected number of tokens and time required for future decoding stages.

For any given request $i$ in the decoding phase, the historical states include the decoding time already spent $T_{\text{past}}^i$ (including time waiting for decoding) and the number of tokens already generated $N_{\text{past}}^i$, while the future entails the estimated number of tokens still required $N_{\text{future}}^i$ and the expected remaining decoding time $T_{\text{future}}^i$.
$T_{\text{past}}^i$ and $N_{\text{past}}^i$ are accessible through direct monitoring, whereas $N_{\text{future}}^i$ and $T_{\text{future}}^i$ require predictive estimation.
Similar to the approach in latest work~\cite{DBLP:journals/corr/abs-2404-08509}, the prediction of the complete generation length $N_{\text{max}}^i$ can be framed as a multi-class classification problem to ensure prediction accuracy. 
Specifically, the predicted complete generation length can be divided into multiple percentile ranges, and a model predicts which range the output sequence length corresponding to the request falls into.
Under this prediction approach, $N_{\text{future}}^i$ is conservatively estimated by subtracting $N_{\text{past}}^i$ from the lower bound of the predicted generation length range.
Naturally, $N_{\text{future}}^i$ is constrained to positive integers.
The expected remaining decoding time $T_{\text{future}}^i$ is simply estimated using the current TPOT multiplied by $N_{\text{future}}^i$.

At this point, for any request in the decoding phase with an SLO target that requires TPOT to be less than $T_{\text{tpot}}^i$ seconds, the maximum allowable duration for scheduling the prefill of new requests, $T_{\text{allow\_prefill}}^i$, can be calculated as follows:
\begin{equation}
\begin{aligned}
T_{\text{allow\_prefill}}^i = T_{\text{tpot}}^i \times \left( N_{\text{past}}^i + N_{\text{future}}^i \right) - \left( T_{\text{past}}^i + T_{\text{future}}^i \right)
\end{aligned}
\label{eq:allow}
\end{equation}

Given the set of requests currently in the Request Queue $\{q_1, q_2, \dots\} \in Q$, the prefill stages for requests from $q_1$ to $q_n$ can be scheduled as long as the following condition is met:
\begin{equation}
\begin{aligned}
\sum_{k=1}^{n} T_{\text{prefill}}^{q_k} < \min_{i} \left( T_{\text{allow\_prefill}}^i \right)
\end{aligned}
\end{equation}

The prefill time $T_{\text{prefill}}$ for each request $q_k$ can be estimated using the following formula:
\begin{equation}
\begin{aligned}
T_{\text{prefill}} = \alpha \times \text{seqlen} \times \frac{2 \times n_{\text{param}} + 2 \times \text{seqlen} \times n_{\text{hidden}}}{\text{FLOP per second of device}}
\end{aligned}
\label{eq:prefill}
\end{equation}
where $\text{seqlen}$ denotes the sequence length of the prompt; $n_{\text{param}}$ and $_{\text{hidden}}$ denote the model’s total number of parameters and hidden layer size, respectively; 
$\alpha$ is an empirical correction factor derived from profiling data that adjusts the theoretical estimate to more accurately reflect the observed prefill times under real-world conditions.

Algorithm~\ref{alg:slo_scheduler} summarizes the process through which the SLO-aware Scheduler makes request scheduling decisions based on TPOT SLOs.
\begin{algorithm}
\small
\caption{Pseudo-code of SLO-aware Scheduler}
\label{alg:slo_scheduler}
\begin{algorithmic}[1]
\Require 
Request Queue $Q = \{q_1, q_2, \dots\}$, Requests in decoding phase $D = \{d_1, d_2, \dots\}$ with TPOT SLO targets;
\For{each $d_i \in D$}
    \State Estimate $T_{\text{allow\_prefill}}^i$ using Eq.~\ref{eq:allow};
\EndFor

\For{each $q_k \in Q$}
    \State Estimate $T_{\text{prefill}}^{q_k}$ using Eq.~\ref{eq:prefill};
\EndFor

\State Initialize $n \leftarrow 0$;
\While{$\sum_{k=1}^{n+1} T_{\text{prefill}}^{q_k} < \min_{i} \left( T_{\text{allow\_prefill}}^i \right)$}
    \State $n \leftarrow n + 1$;
\EndWhile

\State \Return $n$ \Comment{Maximum number of requests for prefill scheduling}
\end{algorithmic}
\end{algorithm}

\subsubsection{Layer-wise KV Blocks Allocation}
Once the SLO-aware Scheduler determines the scheduling of prefill stages for specific requests, these requests are subsequently by the LayerKV Execution Engine.
The processing specifically involves layer-wise KV block allocation and KV cache offloading, enabling the limited GPU KV blocks to support more incoming requests.

A critical consideration in layer-wise KV block allocation is determining the minimum number of layers that must be retained within the GPU KV blocks to ensure that computation time fully overlaps with offloading communication time, thereby maintaining QPS.
Suppose a model consists of $L$ layers. 
For this model’s KV cache, at least $x$ layers are retained on the GPU, while the remaining $L-x$ layers are offloaded to the CPU.
The offloading is performed asynchronously during the computation.
The prefill time exhibits a superlinear relationship with the sequence length, as shown in Eq.~\ref{eq:prefill}, while the offloading time, which scales linearly with sequence length, can be estimated as follows:
\begin{equation}
\begin{aligned}
T_{\text {offload}} = \beta \times \text { seqlen } \times \frac{2(L-x) \times d_{\text {heads }} \times n_{\text {heads }} \times f_{\text {precision }} }{\text { PCIe Bandwidth }}
\end{aligned}
\label{eq:prefill}
\end{equation}
where $d_{\text{heads}}$ is the dimensionality of each attention head, $n_{\text{heads}}$ refers to the number of attention heads, and $f_{\text{precision}}$ specifies the numerical precision format,
$\beta$ is an empirical correction factor.
To fully conceal the PCIe communication overhead, the condition $T_{\text{offload}} \leq T_{\text{prefill}}$ must be satisfied. Based on this condition, the $x$ can be determined. It can be noted that $x$ is closely linked to the length of the requested prompt, as well as the model’s architecture and hardware setup. When the prompt is long, $x$ can be zero, allowing all KV cache layers to be offloaded to the CPU without occupying GPU KV blocks. Conversely, when the prompt is short, $x$ is greater than zero, requiring at least $x$ KV cache layers to remain in GPU memory, as their communication overhead cannot yet be fully overlapped. Note that the minimum number of reserved layers in a GPU KV block does not imply that these KV caches must remain in GPU memory for an extended duration. They also can be offloaded to the CPU, freeing up GPU memory during stages when PCIe are relatively idle. Here, GPU KV blocks can be regarded as a special send buffer.

Certainly, keeping certain layers of the KV cache in the GPU until they are used offers clear advantages, which can be considered free prefetching.
However, if GPU KV resources become insufficient after multiple inference phases, this may block or preempt other requests, adversely affecting system QPS—an outcome we aim to avoid.
Therefore, we introduce a strategy to evaluate whether further offloading of these $x$ layers is required based on system resource availability.
Concretely, we propose a state transition equation to proactively anticipate the status of GPU KV blocks across several stages.
The equation is defined as:
\begin{equation}
\begin{aligned}
\text{Avail}(t+1) = \text{Avail}(t) + \text{Released}(t) - \text{Allocated}(t)
\end{aligned}
\label{eq:memory_transition}
\end{equation}
where $\text{Avail}(t)$ and $\text{Avail}(t+1)$ represent the number of free KV blocks at the beginning of stages $t$ and $t+1$, respectively. 
$\text{Released}(t)$ and $\text{Allocated}(t)$ denote the KV blocks released and allocated at time $t$.
First, the initial value of $\text{Avail}(t)$ can be based on the current number of available GPU KV blocks. 
Second, $\text{Released}(t)$ is defined as the quantity of GPU KV blocks freed by sequences that have concluded at the current time. 
To estimate which sequences will finish, the multi-class prediction model discussed in §~\ref{subsec:slo} is utilized, with the median of the predicted sequence length range serving as a rough estimate.
Third, the estimation of $\text{Allocated}(t)$ considers both the number of sequences at time t and the KV cache scheduled for prefill or decoding. 
We conservatively assume that each sequence requires one additional KV block, and for decoding, any new request in the running queue will be included in the batch if the available blocks are sufficient.
As for the KV blocks required for prefill, are the variables that need to be controlled.

When the available GPU KV blocks fall below the preset threshold, indicating resource insufficiency, the retained $x$ layers of KV cache will be offloaded to the CPU. 
We prioritize offloading the most recently processed requests, starting with $x/2$ layers. 
If this proves insufficient, the full offloading will be executed.

\subsubsection{Layer-wise KV Cache Management}

In the design of LayerKV, layer-based KV cache management is a critical strategy that optimizes system resource utilization by alternating caching of KV layers between the GPU and CPU.
However, to effectively manage the mapping between these KV cache layers and their corresponding devices, an additional table is required to store the mapping information for each KV cache layer and its assigned device.

Specifically, we determine the number of layers to retain on the GPU by considering available device memory and the total token count of the request, which guides the execution of LayerKV offloading. The offloaded layers are evenly distributed across the model’s layers. For example, in an 8-layer model, if 4 layers of KV cache are kept on the GPU, we retain the 1st, 3rd, 5th, and 7th layers on the GPU, while the 0th, 2nd, 4th, and 6th layers are offloaded to the CPU. This approach allows computation to overlap with transmission overhead, as only the KV cache for the 0th layer must be offloaded before the attention operation in the 2nd layer begins, enabling data transfer during the computation of the 0th and 1st layers.
Moreover, since block location information varies between layers, we extend the block table, which records the block ID and storage location for each request.
We add layer-wise information to each block, indicating the indices of the layers where the KV cache is retained on the GPU and the indices of the layers stored on the CPU.

\subsubsection{Layer-wise KV Cache Offloading on Multi-GPUs}
When the model weights of a LLM exceed the capacity of a single GPU, multiple GPUs are typically deployed using tensor parallelism, where both model weights and KV cache are distributed across GPUs.
During the forward pass of each layer, both the computation of Attention and FFN require an all-reduce operation.
On GPU nodes equipped with NVLink, this all-reduce operation transfers data via NVLink, which does not interfere with LayerKV swapping between the CPU and GPU.
However, on GPU nodes without NVLink, the all-reduce operation transfers data over PCIe, which is also used by LayerKV.
This leads to PCIe contention, as the all-reduce operation is on the critical path of end-to-end inference latency and directly impacts system throughput.

To mitigate PCIe contention, LayerKV implements a mechanism that checks PCIe usage before initiating swapping.
If PCIe is already in use, the swapping operation is delayed for a portion of the all-reduce latency before checking again.
This check mechanism ensures that LayerKV swapping is not launched during an ongoing all-reduce operation.
Additionally, to further alleviate contention, the swapping data is divided into smaller units, and the check mechanism is applied to each subunit, reducing interference with the ongoing all-reduce operation.
Together, these methods significantly reduce PCIe contention.

\section{Implementation}
\label{sec:implementation}

\sysname is implemented based on the widely adopted LLM inference framework vLLM~\cite{kwon2023efficient}. 
To ensure the SLO requirements for TTFT and TPOT are met, the scheduler orchestrates batch requests for both the prefill and decode stages. 
Prior to each scheduling event, the runtime records the queuing time, progress, and predicted sequence length for each request. 
This information is then provided to the scheduler to make decisions for each stage.

For KV cache memory management, \sysname allocates a single PyTorch tensor during initialization to store physical KV cache, rather than assigning a separate tensor for each layer. 
This approach allows flexible logical allocation for partial layers within a request, which is essential for layer-wise KV cache management. 
To expedite the transfer of KV cache between the CPU and GPU, a dedicated CUDA stream is introduced, enabling computation and data transfer to occur concurrently. 
Additionally, a CPU thread handles the transfer of KV cache from pinned memory to pageable memory, preventing delays in GPU computation caused by KV transfers and copies.

In the prefill stage, the \textit{h2d} (host-to-device) transfer of KV cache is initiated immediately after KV computation for each layer, overlapping with the computation of the same layer. 
In contrast, during the decode stage, KV cache is transferred layer-by-layer from host memory to GPU memory.

\section{Evaluation}
\label{sec:evaluation}

In this section, we evaluate the performance of \sysname with state-of-the-art solutions on various LLM models with different real-world workloads.
The evaluation shows \sysname outperforms the current state-of-the-art system in terms of TTFT under the same TPOT SLA requirements.

\begin{figure*}[t]
    \includegraphics[width=\linewidth]{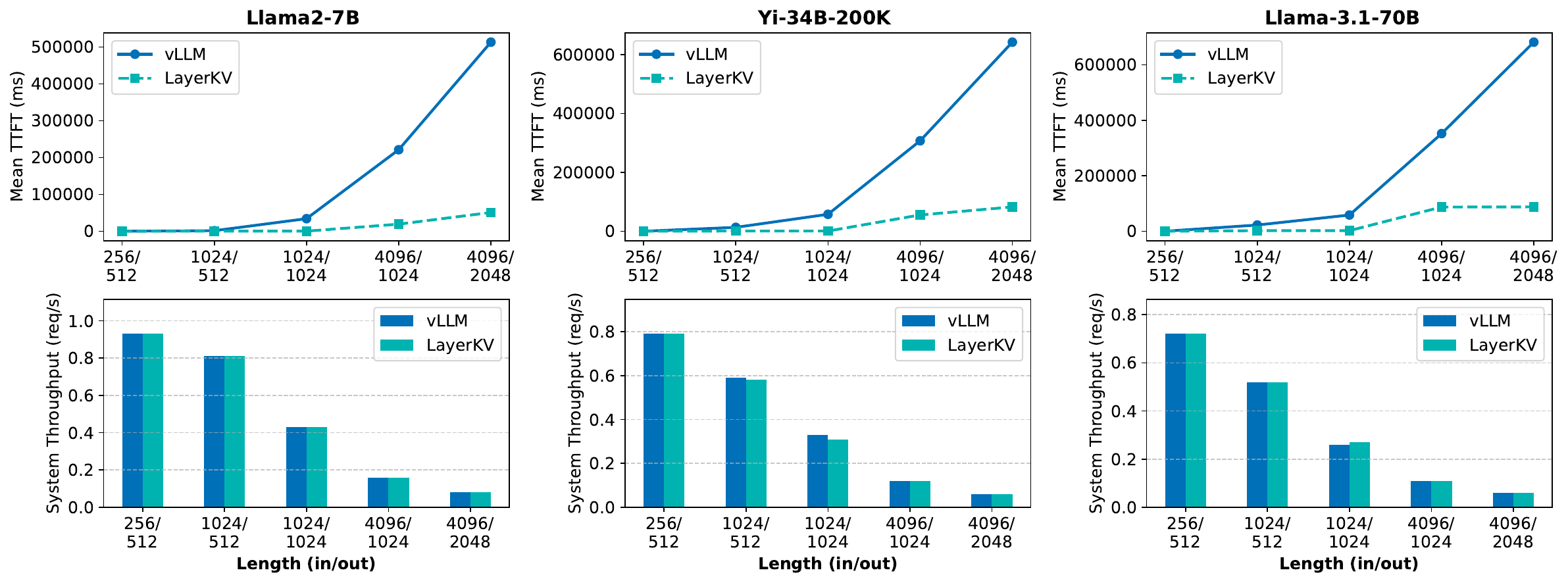}
    \caption{Performance Comparison of LayerKV and vLLM Under Varying Context Lengths.}
    \label{fig:evaluation:Context}
\end{figure*}
\begin{figure*}[t]
    \includegraphics[width=0.66\linewidth]{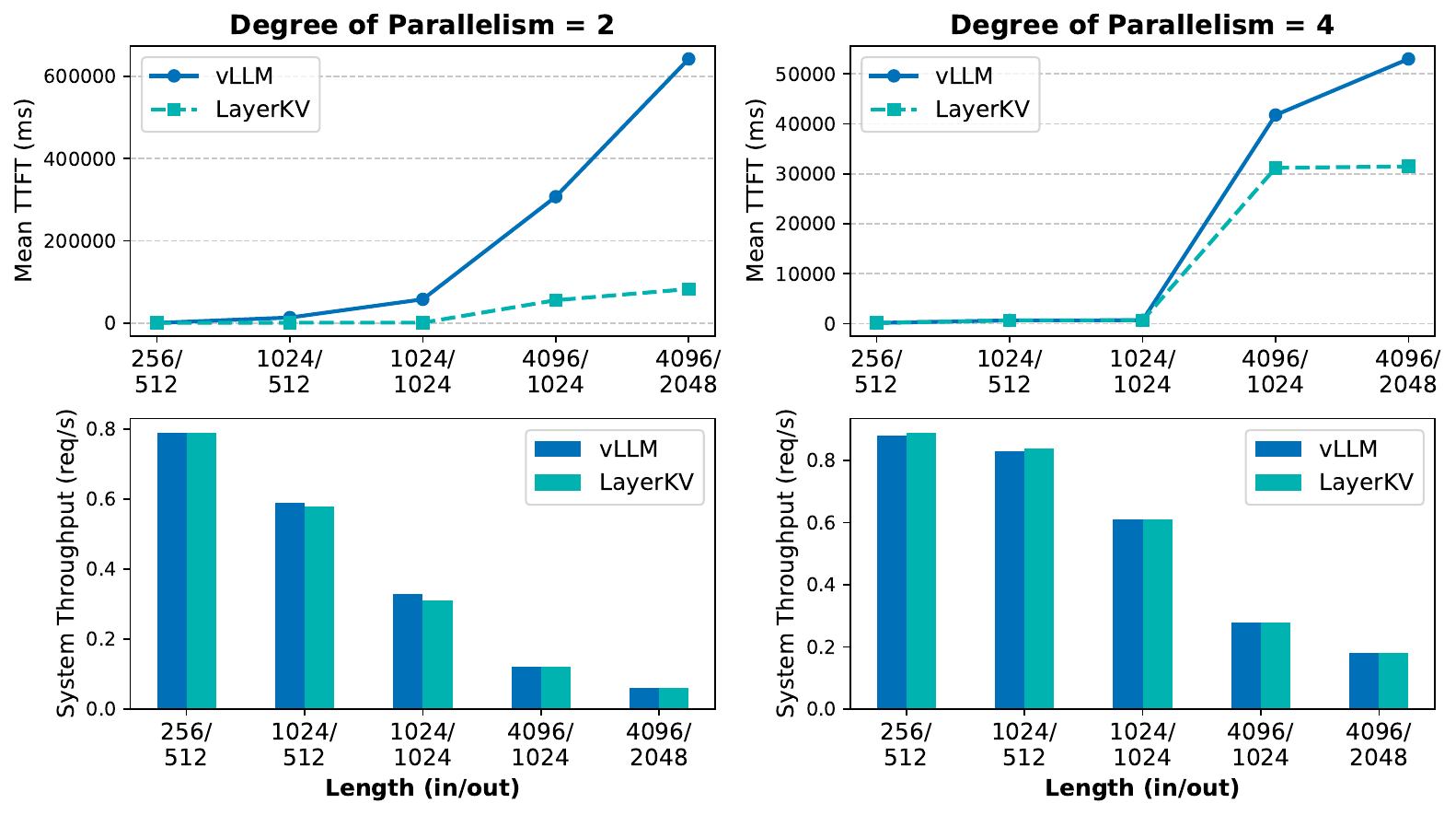}
    \caption{Performance Comparison of LayerKV and vLLM Under Varying Degree of Parallelism.}
    \label{fig:evaluation:DoP}
\end{figure*}

\subsection{Experimental Setup}
\label{sec:evaluation:setup}

\parabf{Models.}
We use \emph{Llama-2-7B}~\cite{touvron2023llama2}, \emph{Yi-34B-200K}~\cite{DBLP:Yi}, \emph{and Llama-3.1-70B} as the LLM model in our evaluation. 
As these models are widely used in academic and industry, and have different model size and the longest request length targeting different application scenarios. 
In addition, Yi-34B-200K and Llama-3.1-70B supports memory efficient attention technique, grouped-query attention (GQA)~\cite{DBLP:GQA}, which saves KV memory footprint.

\parabf{Testbed.}
We evaluate \sysname on servers each with eight NVIDIA L20 48GB GPUs, 64 CPU cores, 2048 GB of host memory. 
The PCIe is used to connect GPUs and CPUs, and each two GPUs share one PCIe connection. 
We use PyTorch 2.4.0, CUDA 12.2, vLLM 0.5.5 for our evaluation. 
All experiments are conducted on this server, with the number of GPUs adjusted based on model requirements: 1 GPU for Llama2-7B, 2 GPUs for Yi-34B, and 4 GPUs for Llama3.1-70B. The degree of tensor parallelism is set to 1, 2, and 4, respectively.

\parabf{Workloads.} 
We employed fixed-length inputs to intuitively demonstrate system performance across different context lengths, while also incorporating a popular real-world dataset, ShareGPT~\cite{sharegpt}, to simulate practical service scenarios. 
The dataset, collected from real conversations with ChatGPT, has been widely utilized in prior research~\cite{kwon2023efficient,wu2023fast,zhong2024distserve}. Due to the limited context window of ChatGPT-3.5, the sequence length in this dataset ranges from 4 to 2.3K tokens.

\parabf{Baselines.}
We compare \sysname with the following state-of-the-art LLM serving systems: \emph{vLLM}~\cite{kwon2023efficient}\footnote{vLLM 0.5.5}: It is one of the most popular LLM serving systems.
\begin{figure*}[t]
    \includegraphics[width=\linewidth]{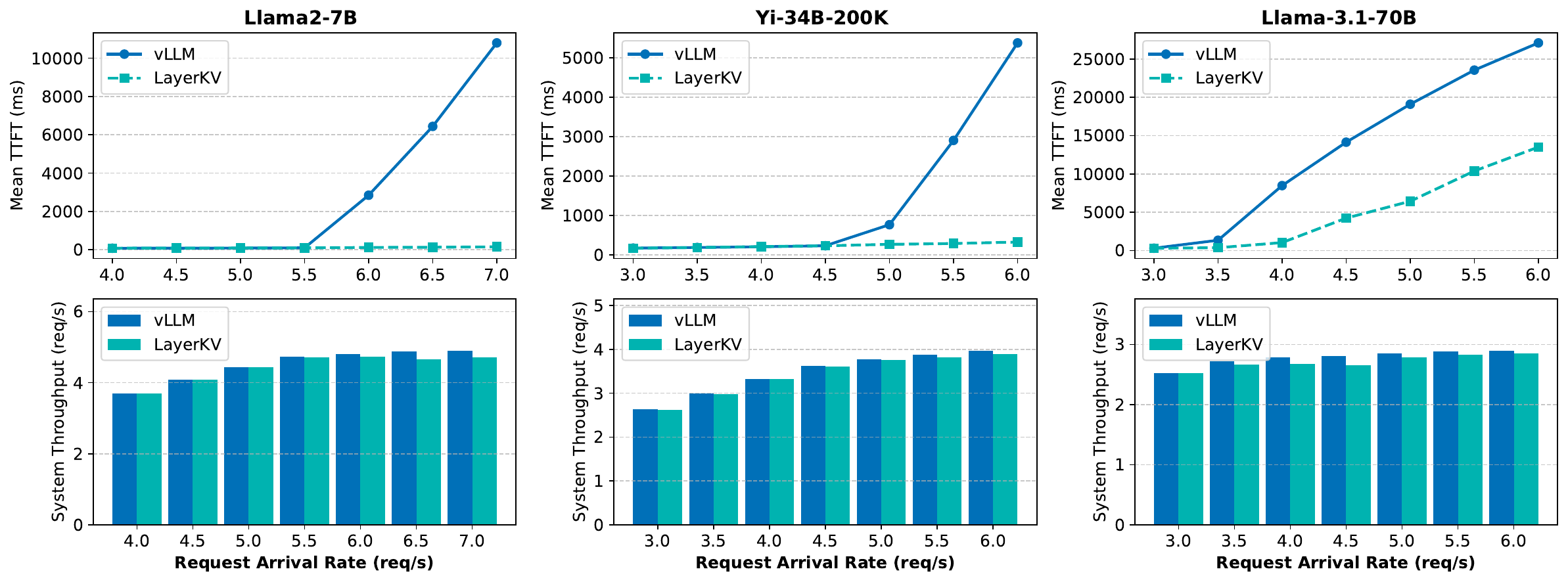}
    \caption{Performance Comparison of LayerKV and vLLM Under Varying Request Arrival Rates.}
    \label{fig:evaluation:QPS}
\end{figure*}
\subsection{End-to-End Performance}
\subsubsection{Performance Comparison Under Varying Context Lengths}

Figure~\ref{fig:evaluation:Context} presents a comparison of performance between LayerKV and vLLM across varying context lengths, with the request arrival rate of 1 req/s.
The top three line plots indicate that as the context length increases, vLLM’s TTFT escalates sharply, while LayerKV experiences a more gradual rise, with the performance gap widening up to an order of magnitude. 
This outcome aligns with LayerKV's core design principles, as previously discussed. 

In contrast, the lower three bar charts show that throughput of both systems naturally decrease with increasing context length.
At this request arrival rate, the throughput of the two systems is nearly identical.

\subsubsection{Performance Comparison Under Varying Degree of Parallelism}

We further investigate the impact of degree of parallelism (DoP) on the Yi-34B-200K model, as illustrated in Figure~\ref{fig:evaluation:DoP}. 
With increasing DoP, computational capacity scales proportionally, and larger GPU memory alleviates resource contention. 
Despite these improvements, LayerKV consistently achieves notable TTFT reductions. 
Additionally, the increased DoP further narrows the marginal throughput gap between LayerKV and vLLM.

\subsubsection{Performance Comparison Under Varying Request Arrival Rates}

\begin{figure}[t]
    \includegraphics[width=\linewidth]{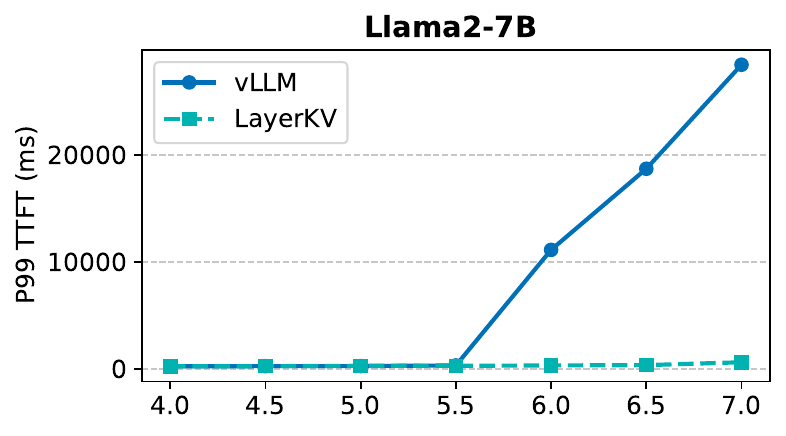}
    \caption{P99 TTFT Comparison of LayerKV and vLLM Under Varying Request Arrival Rates.}
    \label{fig:evaluation:p99}
\end{figure}

Figure~\ref{fig:evaluation:QPS} compares the performance of LayerKV and vLLM across different request arrival rates using the ShareGPT dataset.
As the request arrival rate increases, vLLM’s TTFT rises sharply, particularly at higher rates, where queuing delays lead to significant latency spikes. 
In contrast, LayerKV effectively controls TTFT, maintaining relatively low latency even under heavy loads.
In this case, LayerKV achieves up to an 69x reduction in mean TTFT latency and 45x reduction in P99 TTFT latency (by Figure~\ref{fig:evaluation:p99}).

Furthermore, under low load conditions, the system’s throughput scales proportionally with the increase in request arrival rate.
However, once the arrival rate exceeds a certain threshold, the system enters an overburdened state, causing throughput to reach a bottleneck.
At this time, due to the need to swap portions of the KV cache from the CPU during the decoding phase, LayerKV's throughput is marginally lower than vLLM's. 
However, LayerKV mitigates this by maximizing the number of layers retained on the GPU through layer-wise KV block allocation, limiting the throughput gap consistently maintaining it below 3\%

\subsubsection{SLO Violation Rate Comparison Under Varying Request Arrival Rates}
\begin{figure}[t]
    \includegraphics[width=\linewidth]{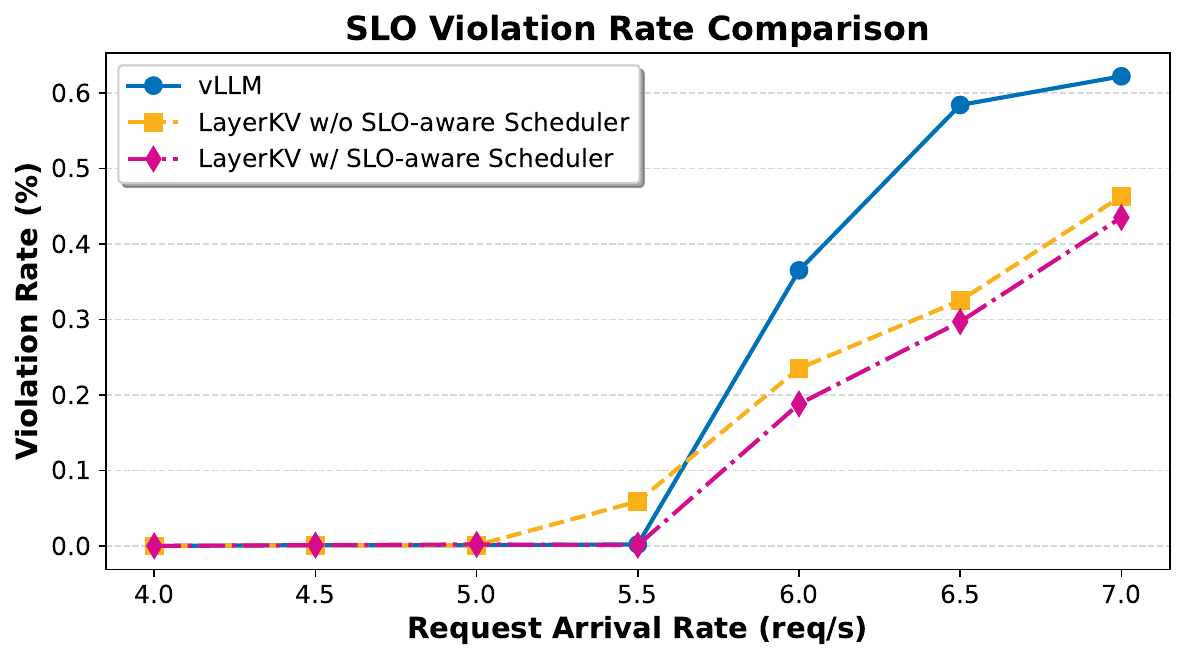}
    \caption{SLO Violation Rate Comparison of LayerKV and vLLM Under Varying Request Arrival Rates.}
    \label{fig:evaluation:vio}
\end{figure}

We further explore the impact of varying request arrival rates on the SLO violation rate using the Llama2-7B model. 
Specifically, the TTFT SLO is set to 3000 ms and the TPOT SLO to 200 ms for each request. 
A violation is recorded if either of these thresholds is exceeded.

Figure~\ref{fig:evaluation:vio} presents the experimental results. 
It is evident that as the request arrival rate reaches 6 requests per second, vLLM begins to exhibit significant SLO violations due to a sharp increase in TTFT. 
LayerKV consistently maintains a violation rate 17.7–28.7\% lower than vLLM.

Furthermore, we observe that without the SLO-aware Scheduler, LayerKV encounters increased TPOT, resulting in some SLO violations. For instance, at a request arrival rate of 5.5, its performance is occasionally inferior to vLLM. However, with the integration of the SLO-aware Scheduler, this issue is effectively mitigated.

\begin{table*}[t!]
    \centering
    \renewcommand{\arraystretch}{1.3} 
    \setlength{\tabcolsep}{10pt} 
    \begin{tabular}{| l | c | c | c |}
        \hline
        \textbf{Inference Framework} & \textbf{KV Cache Management} & \textbf{KV Cache Offloading} & \textbf{SLO-aware Scheduling} \\
        \hline
        vLLM~\cite{kwon2023efficient} & Request-wise & Request-wise & Not support yet \\
        \hline
        DistServe~\cite{zhong2024distserve} & Request-wise & Not support yet & Static \\
        \hline
        DeepSpeed-FastGen~\cite{holmes2024deepspeedfastgen} & Request-wise & Not support yet& Static \\
        \hline
        LayerKV (Ours) & Layer-wise & Layer-wise & Dynamic \\
        \hline
    \end{tabular}
    \vspace{0.4cm}
    \caption{Comparison of LLM Serving Systems}
    \label{tab:compare_frame}
\end{table*}

\section{Related Work}
\subsection{KV Cache Optimization}
In this section, we further explore works related to KV cache memory optimization, which can be broadly categorized into \textbf{algorithm-level optimizations} and \textbf{system-level optimizations}.

\textbf{Algorithm-level optimizations} aim to reduce KV cache memory requirements.
Common techniques include KV quantization~\cite{DBLP:journals/corr/abs-2401-18079,DBLP:conf/icml/LiuYJZXBC024,DBLP:journals/corr/abs-2405-14256}, window attention~\cite{zhang2023h2o,DBLP:streaming-llm}, KV pruning~\cite{DBLP:journals/corr/abs-2407-14057,DBLP:conf/icml/TangZZXKH24}, and activation-shared attention~\cite{DBLP:GQA,DBLP:MQA}. 
These methods seek to lower memory usage by compressing the cache or eliminating unnecessary KV entries, but they often come with a certain degree of accuracy loss.

\textbf{System-level optimizations} focus on increasing available memory capacity. 
A straightforward approach is to utilize multiple GPUs or offload KV caches to CPU memory or even disk when GPU memory is insufficient.
Various parallel strategies partition memory demands across token-wise~\cite{DBLP:journals/corr/abs-2404-09526,DBLP:ring-attention,DBLP:striped-attention}, model-wise~\cite{DBLP:alpa,pipedream}, operator-wise~\cite{DBLP:Megatron-LM,DBLP:conf/mlsys/PopeDCDBHXAD23}, and stage-wise~\cite{DBLP:mooncake,zhong2024distserve} levels, effectively distributing memory pressure.
However, these strategies typically require multi-GPU or distributed environments and are not applicable to single-device setups.
On the other hand, well-established offloading methods~\cite{DBLP:conf/osdi/LeeLSS24,DBLP:conf/icml/0007ZYLRCLRSZ23} are more suitable for offline scenarios, where the primary focus is on optimizing system throughput rather than SLO metrics.
In contrast, LayerKV maintains lossless precision, supports both single- and multi-GPU setups, and simultaneously accounts for SLO requirements.
vTensor~\cite{xu2024vtensor} proposes a new virtual memory abstraction for managing KV cache memory using CUDA virtual memory management, enabling both memory defragmentation and computation flexibility. This optimization is orthogonal to LayerKV.

\subsection{LLM Serving Systems}

Table~\ref{tab:compare_frame} compares LayerKV with state-of-the-art LLM serving systems. 
vLLM~\cite{kwon2023efficient} proposes a PagedAttention mechanism to minimize GPU memory fragmentation, boosting batch size and token generation throughput. 
DistServe~\cite{zhong2024distserve} divides prefill and decode execution to different GPUs in order to avoid performance interference of the two computation stages. 
DeepSpeed-FastGen~\cite{holmes2024deepspeedfastgen} and Sarathi~\cite{agrawal2023sarathi} decompose prompts into small chunks and combine them with decode tokens, which improves responsiveness and tail latency. 

Compared to previous approaches, LayerKV manages the KV cache through a more flexible layer-wise method, expanding the memory management space. 
Layer-wise offloading significantly reduces queuing delays and improves TTFT. 
The SLO-aware scheduling carefully optimizes both TTFT and TPOT SLO.

\section{Conclusion}
To address the significant challenge of increasing Time to First Token (TTFT) in LLM serving under large context lengths, we have developed \sysname, which introduces layer-wise KV block
allocation, management, and offloading for fine-grained control over system memory.
\sysname effectively reduces queuing delays by optimizing GPU KV cache usage without requiring additional hardware or compromising output quality.
Our comprehensive evaluations on models ranging from 7B to 70B parameters across single and multiple GPUs demonstrate that \sysname achieves up to an 69x reduction in TTFT latency and reduces Service Level Objective (SLO) violation rates by 28.7\%, significantly enhancing user experience.

\section{Future Work}
Looking ahead, there are several directions to further enhance the capabilities of \sysname. First, we plan to enhance \sysname by integrating KV cache quantization techniques to further optimize memory efficiency. 
Quantizing KV caches will enable \sysname to support larger models and longer context lengths by reducing resource consumption.
Moreover, we aim to extend \sysname by incorporating prefill-decoding disaggregation to further minimize queuing delays and improve system throughput. 
By decoupling the prefill and decoding stages, we can achieve greater flexibility in resource allocation and more precise adherence to SLO requirements.
These features will be released in future versions.

\label{lastpage}


\bibliographystyle{style/ACM-Reference-Format}
\bibliography{paper}


\begin{thebibliography}{00}


\ifx \showCODEN    \undefined \def \showCODEN     #1{\unskip}     \fi
\ifx \showDOI      \undefined \def \showDOI       #1{#1}\fi
\ifx \showISBNx    \undefined \def \showISBNx     #1{\unskip}     \fi
\ifx \showISBNxiii \undefined \def \showISBNxiii  #1{\unskip}     \fi
\ifx \showISSN     \undefined \def \showISSN      #1{\unskip}     \fi
\ifx \showLCCN     \undefined \def \showLCCN      #1{\unskip}     \fi
\ifx \shownote     \undefined \def \shownote      #1{#1}          \fi
\ifx \showarticletitle \undefined \def \showarticletitle #1{#1}   \fi
\ifx \showURL      \undefined \def \showURL       {\relax}        \fi
\providecommand\bibfield[2]{#2}
\providecommand\bibinfo[2]{#2}
\providecommand\natexlab[1]{#1}
\providecommand\showeprint[2][]{arXiv:#2}

\bibitem[\protect\citeauthoryear{??}{cha}{2022}]%
        {chatgpt}
 \bibinfo{year}{2022}\natexlab{}.
\newblock \bibinfo{title}{Introducing ChatGPT}.
\newblock \bibinfo{howpublished}{\url{https://openai.com/blog/chatgpt}}.
  (\bibinfo{year}{2022}).
\newblock


\bibitem[\protect\citeauthoryear{??}{sha}{2023}]%
        {sharegpt}
 \bibinfo{year}{2023}\natexlab{}.
\newblock \bibinfo{title}{ShareGPT Teams}.
\newblock \bibinfo{howpublished}{\url{https://sharegpt.com/}}.
  (\bibinfo{year}{2023}).
\newblock


\bibitem[\protect\citeauthoryear{Agrawal, Panwar, Mohan, Kwatra, Gulavani, and
  Ramjee}{Agrawal et~al\mbox{.}}{2023}]%
        {agrawal2023sarathi}
\bibfield{author}{\bibinfo{person}{Amey Agrawal}, \bibinfo{person}{Ashish
  Panwar}, \bibinfo{person}{Jayashree Mohan}, \bibinfo{person}{Nipun Kwatra},
  \bibinfo{person}{Bhargav~S. Gulavani}, {and} \bibinfo{person}{Ramachandran
  Ramjee}.} \bibinfo{year}{2023}\natexlab{}.
\newblock \showarticletitle{SARATHI: Efficient LLM Inference by Piggybacking
  Decodes with Chunked Prefills}.
\newblock \bibinfo{journal}{{\em arXiv\/}} (\bibinfo{year}{2023}).
\newblock


\bibitem[\protect\citeauthoryear{Ainslie, Lee{-}Thorp, de~Jong, Zemlyanskiy,
  Lebr{\'{o}}n, and Sanghai}{Ainslie et~al\mbox{.}}{2023}]%
        {DBLP:GQA}
\bibfield{author}{\bibinfo{person}{Joshua Ainslie}, \bibinfo{person}{James
  Lee{-}Thorp}, \bibinfo{person}{Michiel de Jong}, \bibinfo{person}{Yury
  Zemlyanskiy}, \bibinfo{person}{Federico Lebr{\'{o}}n}, {and}
  \bibinfo{person}{Sumit Sanghai}.} \bibinfo{year}{2023}\natexlab{}.
\newblock \showarticletitle{{GQA:} Training Generalized Multi-Query Transformer
  Models from Multi-Head Checkpoints}. In \bibinfo{booktitle}{{\em Proceedings
  of the 2023 Conference on Empirical Methods in Natural Language Processing,
  {EMNLP} 2023, Singapore, December 6-10, 2023}},
  \bibfield{editor}{\bibinfo{person}{Houda Bouamor}, \bibinfo{person}{Juan
  Pino}, {and} \bibinfo{person}{Kalika Bali}} (Eds.).
  \bibinfo{publisher}{Association for Computational Linguistics},
  \bibinfo{pages}{4895--4901}.
\newblock


\bibitem[\protect\citeauthoryear{Anthropic}{Anthropic}{2024}]%
        {claude3}
\bibfield{author}{\bibinfo{person}{Anthropic}.}
  \bibinfo{year}{2024}\natexlab{}.
\newblock \bibinfo{title}{Introducing the next generation of Claude}.
\newblock
  \bibinfo{howpublished}{\url{https://www.anthropic.com/news/claude-3-family}}.
    (\bibinfo{year}{2024}).
\newblock


\bibitem[\protect\citeauthoryear{Brandon, Nrusimha, Qian, Ankner, Jin, Song,
  and Ragan{-}Kelley}{Brandon et~al\mbox{.}}{2023}]%
        {DBLP:striped-attention}
\bibfield{author}{\bibinfo{person}{William Brandon}, \bibinfo{person}{Aniruddha
  Nrusimha}, \bibinfo{person}{Kevin Qian}, \bibinfo{person}{Zachary Ankner},
  \bibinfo{person}{Tian Jin}, \bibinfo{person}{Zhiye Song}, {and}
  \bibinfo{person}{Jonathan Ragan{-}Kelley}.} \bibinfo{year}{2023}\natexlab{}.
\newblock \showarticletitle{Striped Attention: Faster Ring Attention for Causal
  Transformers}.
\newblock \bibinfo{journal}{{\em CoRR\/}}  \bibinfo{volume}{abs/2311.09431}
  (\bibinfo{year}{2023}).
\newblock


\bibitem[\protect\citeauthoryear{Frantar, Ashkboos, Hoefler, and
  Alistarh}{Frantar et~al\mbox{.}}{2022}]%
        {DBLP:gptq}
\bibfield{author}{\bibinfo{person}{Elias Frantar}, \bibinfo{person}{Saleh
  Ashkboos}, \bibinfo{person}{Torsten Hoefler}, {and} \bibinfo{person}{Dan
  Alistarh}.} \bibinfo{year}{2022}\natexlab{}.
\newblock \showarticletitle{{GPTQ:} Accurate Post-Training Quantization for
  Generative Pre-trained Transformers}.
\newblock \bibinfo{journal}{{\em CoRR\/}}  \bibinfo{volume}{abs/2210.17323}
  (\bibinfo{year}{2022}).
\newblock


\bibitem[\protect\citeauthoryear{Fu, Cho, Merth, Mehta, Rastegari, and
  Najibi}{Fu et~al\mbox{.}}{2024}]%
        {DBLP:journals/corr/abs-2407-14057}
\bibfield{author}{\bibinfo{person}{Qichen Fu}, \bibinfo{person}{Minsik Cho},
  \bibinfo{person}{Thomas Merth}, \bibinfo{person}{Sachin Mehta},
  \bibinfo{person}{Mohammad Rastegari}, {and} \bibinfo{person}{Mahyar Najibi}.}
  \bibinfo{year}{2024}\natexlab{}.
\newblock \showarticletitle{LazyLLM: Dynamic Token Pruning for Efficient Long
  Context {LLM} Inference}.
\newblock \bibinfo{journal}{{\em CoRR\/}}  \bibinfo{volume}{abs/2407.14057}
  (\bibinfo{year}{2024}).
\newblock


\bibitem[\protect\citeauthoryear{Gao, Yu, Wu, and Li}{Gao
  et~al\mbox{.}}{2018}]%
        {DBLP:batchmaker}
\bibfield{author}{\bibinfo{person}{Pin Gao}, \bibinfo{person}{Lingfan Yu},
  \bibinfo{person}{Yongwei Wu}, {and} \bibinfo{person}{Jinyang Li}.}
  \bibinfo{year}{2018}\natexlab{}.
\newblock \showarticletitle{Low latency {RNN} inference with cellular
  batching}. In \bibinfo{booktitle}{{\em Proceedings of the Thirteenth EuroSys
  Conference, EuroSys 2018, Porto, Portugal, April 23-26, 2018}},
  \bibfield{editor}{\bibinfo{person}{Rui Oliveira}, \bibinfo{person}{Pascal
  Felber}, {and} \bibinfo{person}{Y.~Charlie Hu}} (Eds.).
  \bibinfo{publisher}{{ACM}}, \bibinfo{pages}{31:1--31:15}.
\newblock


\bibitem[\protect\citeauthoryear{Google}{Google}{2024}]%
        {gemini}
\bibfield{author}{\bibinfo{person}{Google}.} \bibinfo{year}{2024}\natexlab{}.
\newblock \bibinfo{title}{Our next-generation model: Gemini 1.5}.
\newblock
  \bibinfo{howpublished}{\url{https://blog.google/technology/ai/google-gemini-next-generation-model-february-2024/}}.
    (\bibinfo{year}{2024}).
\newblock


\bibitem[\protect\citeauthoryear{Gravitas}{Gravitas}{2023}]%
        {significantgravitasautoGPT}
\bibfield{author}{\bibinfo{person}{Significant Gravitas}.}
  \bibinfo{year}{2023}\natexlab{}.
\newblock \bibinfo{title}{AutoGPT}.
\newblock   (\bibinfo{year}{2023}).
\newblock
\showURL{%
\url{https://github.com/Significant-Gravitas/AutoGPT}}


\bibitem[\protect\citeauthoryear{Han, Zhang, Chen, and Chen}{Han
  et~al\mbox{.}}{2022}]%
        {DBLP:conf/osdi/HanZ0022}
\bibfield{author}{\bibinfo{person}{Mingcong Han}, \bibinfo{person}{Hanze
  Zhang}, \bibinfo{person}{Rong Chen}, {and} \bibinfo{person}{Haibo Chen}.}
  \bibinfo{year}{2022}\natexlab{}.
\newblock \showarticletitle{Microsecond-scale Preemption for Concurrent
  GPU-accelerated {DNN} Inferences}. In \bibinfo{booktitle}{{\em 16th {USENIX}
  Symposium on Operating Systems Design and Implementation, {OSDI} 2022,
  Carlsbad, CA, USA, July 11-13, 2022}},
  \bibfield{editor}{\bibinfo{person}{Marcos~K. Aguilera} {and}
  \bibinfo{person}{Hakim Weatherspoon}} (Eds.). \bibinfo{publisher}{{USENIX}
  Association}, \bibinfo{pages}{539--558}.
\newblock


\bibitem[\protect\citeauthoryear{He, Zhang, Wu, Liu, Zhou, and Zhuang}{He
  et~al\mbox{.}}{2024}]%
        {DBLP:journals/corr/abs-2405-14256}
\bibfield{author}{\bibinfo{person}{Yefei He}, \bibinfo{person}{Luoming Zhang},
  \bibinfo{person}{Weijia Wu}, \bibinfo{person}{Jing Liu},
  \bibinfo{person}{Hong Zhou}, {and} \bibinfo{person}{Bohan Zhuang}.}
  \bibinfo{year}{2024}\natexlab{}.
\newblock \showarticletitle{ZipCache: Accurate and Efficient {KV} Cache
  Quantization with Salient Token Identification}.
\newblock \bibinfo{journal}{{\em CoRR\/}}  \bibinfo{volume}{abs/2405.14256}
  (\bibinfo{year}{2024}).
\newblock


\bibitem[\protect\citeauthoryear{Holmes, Tanaka, Wyatt, Awan, Rasley,
  Rajbhandari, Aminabadi, Qin, Bakhtiari, Kurilenko, and He}{Holmes
  et~al\mbox{.}}{2024}]%
        {holmes2024deepspeedfastgen}
\bibfield{author}{\bibinfo{person}{Connor Holmes}, \bibinfo{person}{Masahiro
  Tanaka}, \bibinfo{person}{Michael Wyatt}, \bibinfo{person}{Ammar~Ahmad Awan},
  \bibinfo{person}{Jeff Rasley}, \bibinfo{person}{Samyam Rajbhandari},
  \bibinfo{person}{Reza~Yazdani Aminabadi}, \bibinfo{person}{Heyang Qin},
  \bibinfo{person}{Arash Bakhtiari}, \bibinfo{person}{Lev Kurilenko}, {and}
  \bibinfo{person}{Yuxiong He}.} \bibinfo{year}{2024}\natexlab{}.
\newblock \showarticletitle{DeepSpeed-FastGen: High-throughput Text Generation
  for LLMs via MII and DeepSpeed-Inference}.
\newblock \bibinfo{journal}{{\em arXiv\/}} (\bibinfo{year}{2024}).
\newblock


\bibitem[\protect\citeauthoryear{Hooper, Kim, Mohammadzadeh, Mahoney, Shao,
  Keutzer, and Gholami}{Hooper et~al\mbox{.}}{2024}]%
        {DBLP:journals/corr/abs-2401-18079}
\bibfield{author}{\bibinfo{person}{Coleman Hooper}, \bibinfo{person}{Sehoon
  Kim}, \bibinfo{person}{Hiva Mohammadzadeh}, \bibinfo{person}{Michael~W.
  Mahoney}, \bibinfo{person}{Yakun~Sophia Shao}, \bibinfo{person}{Kurt
  Keutzer}, {and} \bibinfo{person}{Amir Gholami}.}
  \bibinfo{year}{2024}\natexlab{}.
\newblock \showarticletitle{KVQuant: Towards 10 Million Context Length {LLM}
  Inference with {KV} Cache Quantization}.
\newblock \bibinfo{journal}{{\em CoRR\/}}  \bibinfo{volume}{abs/2401.18079}
  (\bibinfo{year}{2024}).
\newblock


\bibitem[\protect\citeauthoryear{Jiang, Li, Zhang, Wu, Luo, Ahn, Han, Abdi, Li,
  Lin, Yang, and Qiu}{Jiang et~al\mbox{.}}{2024}]%
        {DBLP:MInference}
\bibfield{author}{\bibinfo{person}{Huiqiang Jiang}, \bibinfo{person}{Yucheng
  Li}, \bibinfo{person}{Chengruidong Zhang}, \bibinfo{person}{Qianhui Wu},
  \bibinfo{person}{Xufang Luo}, \bibinfo{person}{Surin Ahn},
  \bibinfo{person}{Zhenhua Han}, \bibinfo{person}{Amir~H. Abdi},
  \bibinfo{person}{Dongsheng Li}, \bibinfo{person}{Chin{-}Yew Lin},
  \bibinfo{person}{Yuqing Yang}, {and} \bibinfo{person}{Lili Qiu}.}
  \bibinfo{year}{2024}\natexlab{}.
\newblock \showarticletitle{MInference 1.0: Accelerating Pre-filling for
  Long-Context LLMs via Dynamic Sparse Attention}.
\newblock \bibinfo{journal}{{\em CoRR\/}}  \bibinfo{volume}{abs/2407.02490}
  (\bibinfo{year}{2024}).
\newblock


\bibitem[\protect\citeauthoryear{Katharopoulos, Vyas, Pappas, and
  Fleuret}{Katharopoulos et~al\mbox{.}}{2020}]%
        {DBLP:Linear-Attention}
\bibfield{author}{\bibinfo{person}{Angelos Katharopoulos},
  \bibinfo{person}{Apoorv Vyas}, \bibinfo{person}{Nikolaos Pappas}, {and}
  \bibinfo{person}{Fran{\c{c}}ois Fleuret}.} \bibinfo{year}{2020}\natexlab{}.
\newblock \showarticletitle{Transformers are RNNs: Fast Autoregressive
  Transformers with Linear Attention}. In \bibinfo{booktitle}{{\em Proceedings
  of the 37th International Conference on Machine Learning, {ICML} 2020, 13-18
  July 2020, Virtual Event}} {\em (\bibinfo{series}{Proceedings of Machine
  Learning Research})}, Vol.~\bibinfo{volume}{119}.
  \bibinfo{publisher}{{PMLR}}, \bibinfo{pages}{5156--5165}.
\newblock


\bibitem[\protect\citeauthoryear{Kwon, Li, Zhuang, Sheng, Zheng, Yu, Gonzalez,
  Zhang, and Stoica}{Kwon et~al\mbox{.}}{2023}]%
        {kwon2023efficient}
\bibfield{author}{\bibinfo{person}{Woosuk Kwon}, \bibinfo{person}{Zhuohan Li},
  \bibinfo{person}{Siyuan Zhuang}, \bibinfo{person}{Ying Sheng},
  \bibinfo{person}{Lianmin Zheng}, \bibinfo{person}{Cody~Hao Yu},
  \bibinfo{person}{Joseph Gonzalez}, \bibinfo{person}{Hao Zhang}, {and}
  \bibinfo{person}{Ion Stoica}.} \bibinfo{year}{2023}\natexlab{}.
\newblock \showarticletitle{Efficient memory management for large language
  model serving with pagedattention}. In \bibinfo{booktitle}{{\em ACM SOSP}}.
\newblock


\bibitem[\protect\citeauthoryear{Lee, Lee, Seo, and Sim}{Lee
  et~al\mbox{.}}{2024}]%
        {DBLP:conf/osdi/LeeLSS24}
\bibfield{author}{\bibinfo{person}{Wonbeom Lee}, \bibinfo{person}{Jungi Lee},
  \bibinfo{person}{Junghwan Seo}, {and} \bibinfo{person}{Jaewoong Sim}.}
  \bibinfo{year}{2024}\natexlab{}.
\newblock \showarticletitle{InfiniGen: Efficient Generative Inference of Large
  Language Models with Dynamic {KV} Cache Management}. In
  \bibinfo{booktitle}{{\em 18th {USENIX} Symposium on Operating Systems Design
  and Implementation, {OSDI} 2024, Santa Clara, CA, USA, July 10-12, 2024}},
  \bibfield{editor}{\bibinfo{person}{Ada Gavrilovska} {and}
  \bibinfo{person}{Douglas~B. Terry}} (Eds.). \bibinfo{publisher}{{USENIX}
  Association}, \bibinfo{pages}{155--172}.
\newblock


\bibitem[\protect\citeauthoryear{Lin, Tang, Tang, Yang, Chen, Wang, Xiao, Dang,
  Gan, and Han}{Lin et~al\mbox{.}}{2024}]%
        {DBLP:awq}
\bibfield{author}{\bibinfo{person}{Ji Lin}, \bibinfo{person}{Jiaming Tang},
  \bibinfo{person}{Haotian Tang}, \bibinfo{person}{Shang Yang},
  \bibinfo{person}{Wei{-}Ming Chen}, \bibinfo{person}{Wei{-}Chen Wang},
  \bibinfo{person}{Guangxuan Xiao}, \bibinfo{person}{Xingyu Dang},
  \bibinfo{person}{Chuang Gan}, {and} \bibinfo{person}{Song Han}.}
  \bibinfo{year}{2024}\natexlab{}.
\newblock \showarticletitle{{AWQ:} Activation-aware Weight Quantization for
  On-Device {LLM} Compression and Acceleration}. In \bibinfo{booktitle}{{\em
  Proceedings of the Seventh Annual Conference on Machine Learning and Systems,
  MLSys 2024, Santa Clara, CA, USA, May 13-16, 2024}},
  \bibfield{editor}{\bibinfo{person}{Phillip~B. Gibbons},
  \bibinfo{person}{Gennady Pekhimenko}, {and} \bibinfo{person}{Christopher~De
  Sa}} (Eds.). \bibinfo{publisher}{mlsys.org}.
\newblock


\bibitem[\protect\citeauthoryear{Liu, Yan, Zaharia, and Abbeel}{Liu
  et~al\mbox{.}}{2024a}]%
        {liu2023world}
\bibfield{author}{\bibinfo{person}{Hao Liu}, \bibinfo{person}{Wilson Yan},
  \bibinfo{person}{Matei Zaharia}, {and} \bibinfo{person}{Pieter Abbeel}.}
  \bibinfo{year}{2024}\natexlab{a}.
\newblock \showarticletitle{World Model on Million-Length Video and Language
  with RingAttention}.
\newblock \bibinfo{journal}{{\em arXiv\/}} (\bibinfo{year}{2024}).
\newblock


\bibitem[\protect\citeauthoryear{Liu, Zaharia, and Abbeel}{Liu
  et~al\mbox{.}}{2023}]%
        {DBLP:ring-attention}
\bibfield{author}{\bibinfo{person}{Hao Liu}, \bibinfo{person}{Matei Zaharia},
  {and} \bibinfo{person}{Pieter Abbeel}.} \bibinfo{year}{2023}\natexlab{}.
\newblock \showarticletitle{Ring Attention with Blockwise Transformers for
  Near-Infinite Context}.
\newblock \bibinfo{journal}{{\em CoRR\/}}  \bibinfo{volume}{abs/2310.01889}
  (\bibinfo{year}{2023}).
\newblock


\bibitem[\protect\citeauthoryear{Liu, Wang, Dao, Zhou, Yuan, Song, Shrivastava,
  Zhang, Tian, R{\'{e}}, and Chen}{Liu et~al\mbox{.}}{2023}]%
        {DBLP:dejavu}
\bibfield{author}{\bibinfo{person}{Zichang Liu}, \bibinfo{person}{Jue Wang},
  \bibinfo{person}{Tri Dao}, \bibinfo{person}{Tianyi Zhou},
  \bibinfo{person}{Binhang Yuan}, \bibinfo{person}{Zhao Song},
  \bibinfo{person}{Anshumali Shrivastava}, \bibinfo{person}{Ce Zhang},
  \bibinfo{person}{Yuandong Tian}, \bibinfo{person}{Christopher R{\'{e}}},
  {and} \bibinfo{person}{Beidi Chen}.} \bibinfo{year}{2023}\natexlab{}.
\newblock \showarticletitle{Deja Vu: Contextual Sparsity for Efficient LLMs at
  Inference Time}. In \bibinfo{booktitle}{{\em International Conference on
  Machine Learning, {ICML} 2023, 23-29 July 2023, Honolulu, Hawaii, {USA}}}
  {\em (\bibinfo{series}{Proceedings of Machine Learning Research})},
  \bibfield{editor}{\bibinfo{person}{Andreas Krause}, \bibinfo{person}{Emma
  Brunskill}, \bibinfo{person}{Kyunghyun Cho}, \bibinfo{person}{Barbara
  Engelhardt}, \bibinfo{person}{Sivan Sabato}, {and} \bibinfo{person}{Jonathan
  Scarlett}} (Eds.), Vol.~\bibinfo{volume}{202}. \bibinfo{publisher}{{PMLR}},
  \bibinfo{pages}{22137--22176}.
\newblock


\bibitem[\protect\citeauthoryear{Liu, Yuan, Jin, Zhong, Xu, Braverman, Chen,
  and Hu}{Liu et~al\mbox{.}}{2024b}]%
        {DBLP:conf/icml/LiuYJZXBC024}
\bibfield{author}{\bibinfo{person}{Zirui Liu}, \bibinfo{person}{Jiayi Yuan},
  \bibinfo{person}{Hongye Jin}, \bibinfo{person}{Shaochen Zhong},
  \bibinfo{person}{Zhaozhuo Xu}, \bibinfo{person}{Vladimir Braverman},
  \bibinfo{person}{Beidi Chen}, {and} \bibinfo{person}{Xia Hu}.}
  \bibinfo{year}{2024}\natexlab{b}.
\newblock \showarticletitle{{KIVI:} {A} Tuning-Free Asymmetric 2bit
  Quantization for {KV} Cache}. In \bibinfo{booktitle}{{\em Forty-first
  International Conference on Machine Learning, {ICML} 2024, Vienna, Austria,
  July 21-27, 2024}}. \bibinfo{publisher}{OpenReview.net}.
\newblock


\bibitem[\protect\citeauthoryear{Narayanan, Harlap, Phanishayee, Seshadri,
  Devanur, Ganger, Gibbons, and Zaharia}{Narayanan et~al\mbox{.}}{2019}]%
        {pipedream}
\bibfield{author}{\bibinfo{person}{Deepak Narayanan}, \bibinfo{person}{Aaron
  Harlap}, \bibinfo{person}{Amar Phanishayee}, \bibinfo{person}{Vivek
  Seshadri}, \bibinfo{person}{Nikhil~R. Devanur}, \bibinfo{person}{Gregory~R.
  Ganger}, \bibinfo{person}{Phillip~B. Gibbons}, {and} \bibinfo{person}{Matei
  Zaharia}.} \bibinfo{year}{2019}\natexlab{}.
\newblock \showarticletitle{PipeDream: Generalized Pipeline Parallelism for DNN
  Training}. In \bibinfo{booktitle}{{\em ACM SOSP}}.
\newblock


\bibitem[\protect\citeauthoryear{Nijkamp, Pang, Hayashi, Tu, Wang, Zhou,
  Savarese, and Xiong}{Nijkamp et~al\mbox{.}}{2023}]%
        {nijkamp2023codegen}
\bibfield{author}{\bibinfo{person}{Erik Nijkamp}, \bibinfo{person}{Bo Pang},
  \bibinfo{person}{Hiroaki Hayashi}, \bibinfo{person}{Lifu Tu},
  \bibinfo{person}{Huan Wang}, \bibinfo{person}{Yingbo Zhou},
  \bibinfo{person}{Silvio Savarese}, {and} \bibinfo{person}{Caiming Xiong}.}
  \bibinfo{year}{2023}\natexlab{}.
\newblock \showarticletitle{CodeGen: An Open Large Language Model for Code with
  Multi-Turn Program Synthesis}. In \bibinfo{booktitle}{{\em International
  Conference on Learning Representations (ICLR)}}.
\newblock


\bibitem[\protect\citeauthoryear{NVIDIA}{NVIDIA}{2019}]%
        {nvidia2019fastertransformer}
\bibfield{author}{\bibinfo{person}{NVIDIA}.} \bibinfo{year}{2019}\natexlab{}.
\newblock \bibinfo{title}{FasterTransformer V1, a highly optimized BERT
  equivalent Transformer layer for inference}.
\newblock
  \bibinfo{howpublished}{\url{https://github.com/NVIDIA/DeepLearningExamples/tree/master/FasterTransformer}}.
    (\bibinfo{year}{2019}).
\newblock
\newblock
\shownote{[Online; accessed April-2020].}


\bibitem[\protect\citeauthoryear{OpenAI}{OpenAI}{2023}]%
        {openai2023gpt4}
\bibfield{author}{\bibinfo{person}{OpenAI}.} \bibinfo{year}{2023}\natexlab{}.
\newblock \bibinfo{title}{GPT-4 Technical Report}.
\newblock   (\bibinfo{year}{2023}).
\newblock


\bibitem[\protect\citeauthoryear{Pope, Douglas, Chowdhery, Devlin, Bradbury,
  Heek, Xiao, Agrawal, and Dean}{Pope et~al\mbox{.}}{2023}]%
        {DBLP:conf/mlsys/PopeDCDBHXAD23}
\bibfield{author}{\bibinfo{person}{Reiner Pope}, \bibinfo{person}{Sholto
  Douglas}, \bibinfo{person}{Aakanksha Chowdhery}, \bibinfo{person}{Jacob
  Devlin}, \bibinfo{person}{James Bradbury}, \bibinfo{person}{Jonathan Heek},
  \bibinfo{person}{Kefan Xiao}, \bibinfo{person}{Shivani Agrawal}, {and}
  \bibinfo{person}{Jeff Dean}.} \bibinfo{year}{2023}\natexlab{}.
\newblock \showarticletitle{Efficiently Scaling Transformer Inference}. In
  \bibinfo{booktitle}{{\em Proceedings of the Sixth Conference on Machine
  Learning and Systems, MLSys 2023, Miami, FL, USA, June 4-8, 2023}},
  \bibfield{editor}{\bibinfo{person}{Dawn Song}, \bibinfo{person}{Michael
  Carbin}, {and} \bibinfo{person}{Tianqi Chen}} (Eds.).
  \bibinfo{publisher}{mlsys.org}.
\newblock


\bibitem[\protect\citeauthoryear{Qin, Li, He, Zhang, Wu, Zheng, and Xu}{Qin
  et~al\mbox{.}}{2024}]%
        {DBLP:mooncake}
\bibfield{author}{\bibinfo{person}{Ruoyu Qin}, \bibinfo{person}{Zheming Li},
  \bibinfo{person}{Weiran He}, \bibinfo{person}{Mingxing Zhang},
  \bibinfo{person}{Yongwei Wu}, \bibinfo{person}{Weimin Zheng}, {and}
  \bibinfo{person}{Xinran Xu}.} \bibinfo{year}{2024}\natexlab{}.
\newblock \showarticletitle{Mooncake: {A} KVCache-centric Disaggregated
  Architecture for {LLM} Serving}.
\newblock \bibinfo{journal}{{\em CoRR\/}}  \bibinfo{volume}{abs/2407.00079}
  (\bibinfo{year}{2024}).
\newblock


\bibitem[\protect\citeauthoryear{Qiu, Mao, Patke, Cui, Jha, Wang, Franke,
  Kalbarczyk, Basar, and Iyer}{Qiu et~al\mbox{.}}{2024}]%
        {DBLP:journals/corr/abs-2404-08509}
\bibfield{author}{\bibinfo{person}{Haoran Qiu}, \bibinfo{person}{Weichao Mao},
  \bibinfo{person}{Archit Patke}, \bibinfo{person}{Shengkun Cui},
  \bibinfo{person}{Saurabh Jha}, \bibinfo{person}{Chen Wang},
  \bibinfo{person}{Hubertus Franke}, \bibinfo{person}{Zbigniew~T. Kalbarczyk},
  \bibinfo{person}{Tamer Basar}, {and} \bibinfo{person}{Ravishankar~K. Iyer}.}
  \bibinfo{year}{2024}\natexlab{}.
\newblock \showarticletitle{Efficient Interactive {LLM} Serving with Proxy
  Model-based Sequence Length Prediction}.
\newblock \bibinfo{journal}{{\em CoRR\/}}  \bibinfo{volume}{abs/2404.08509}
  (\bibinfo{year}{2024}).
\newblock


\bibitem[\protect\citeauthoryear{Shazeer}{Shazeer}{2019}]%
        {DBLP:MQA}
\bibfield{author}{\bibinfo{person}{Noam Shazeer}.}
  \bibinfo{year}{2019}\natexlab{}.
\newblock \showarticletitle{Fast Transformer Decoding: One Write-Head is All
  You Need}.
\newblock \bibinfo{journal}{{\em CoRR\/}}  \bibinfo{volume}{abs/1911.02150}
  (\bibinfo{year}{2019}).
\newblock


\bibitem[\protect\citeauthoryear{Sheng, Zheng, Yuan, Li, Ryabinin, Chen, Liang,
  R{\'{e}}, Stoica, and Zhang}{Sheng et~al\mbox{.}}{2023}]%
        {DBLP:conf/icml/0007ZYLRCLRSZ23}
\bibfield{author}{\bibinfo{person}{Ying Sheng}, \bibinfo{person}{Lianmin
  Zheng}, \bibinfo{person}{Binhang Yuan}, \bibinfo{person}{Zhuohan Li},
  \bibinfo{person}{Max Ryabinin}, \bibinfo{person}{Beidi Chen},
  \bibinfo{person}{Percy Liang}, \bibinfo{person}{Christopher R{\'{e}}},
  \bibinfo{person}{Ion Stoica}, {and} \bibinfo{person}{Ce Zhang}.}
  \bibinfo{year}{2023}\natexlab{}.
\newblock \showarticletitle{FlexGen: High-Throughput Generative Inference of
  Large Language Models with a Single {GPU}}. In \bibinfo{booktitle}{{\em
  International Conference on Machine Learning, {ICML} 2023, 23-29 July 2023,
  Honolulu, Hawaii, {USA}}} {\em (\bibinfo{series}{Proceedings of Machine
  Learning Research})}, \bibfield{editor}{\bibinfo{person}{Andreas Krause},
  \bibinfo{person}{Emma Brunskill}, \bibinfo{person}{Kyunghyun Cho},
  \bibinfo{person}{Barbara Engelhardt}, \bibinfo{person}{Sivan Sabato}, {and}
  \bibinfo{person}{Jonathan Scarlett}} (Eds.), Vol.~\bibinfo{volume}{202}.
  \bibinfo{publisher}{{PMLR}}, \bibinfo{pages}{31094--31116}.
\newblock


\bibitem[\protect\citeauthoryear{Shoeybi, Patwary, Puri, LeGresley, Casper, and
  Catanzaro}{Shoeybi et~al\mbox{.}}{2019}]%
        {DBLP:Megatron-LM}
\bibfield{author}{\bibinfo{person}{Mohammad Shoeybi}, \bibinfo{person}{Mostofa
  Patwary}, \bibinfo{person}{Raul Puri}, \bibinfo{person}{Patrick LeGresley},
  \bibinfo{person}{Jared Casper}, {and} \bibinfo{person}{Bryan Catanzaro}.}
  \bibinfo{year}{2019}\natexlab{}.
\newblock \showarticletitle{Megatron-LM: Training Multi-Billion Parameter
  Language Models Using Model Parallelism}.
\newblock \bibinfo{journal}{{\em CoRR\/}}  \bibinfo{volume}{abs/1909.08053}
  (\bibinfo{year}{2019}).
\newblock


\bibitem[\protect\citeauthoryear{Tang, Zhao, Zhu, Xiao, Kasikci, and Han}{Tang
  et~al\mbox{.}}{2024}]%
        {DBLP:conf/icml/TangZZXKH24}
\bibfield{author}{\bibinfo{person}{Jiaming Tang}, \bibinfo{person}{Yilong
  Zhao}, \bibinfo{person}{Kan Zhu}, \bibinfo{person}{Guangxuan Xiao},
  \bibinfo{person}{Baris Kasikci}, {and} \bibinfo{person}{Song Han}.}
  \bibinfo{year}{2024}\natexlab{}.
\newblock \showarticletitle{{QUEST:} Query-Aware Sparsity for Efficient
  Long-Context {LLM} Inference}. In \bibinfo{booktitle}{{\em Forty-first
  International Conference on Machine Learning, {ICML} 2024, Vienna, Austria,
  July 21-27, 2024}}. \bibinfo{publisher}{OpenReview.net}.
\newblock


\bibitem[\protect\citeauthoryear{Touvron, Martin, Stone, Albert, Almahairi,
  Babaei, Bashlykov, Batra, Bhargava, Bhosale, et~al\mbox{.}}{Touvron
  et~al\mbox{.}}{2023b}]%
        {touvron2023llama2}
\bibfield{author}{\bibinfo{person}{Hugo Touvron}, \bibinfo{person}{Louis
  Martin}, \bibinfo{person}{Kevin Stone}, \bibinfo{person}{Peter Albert},
  \bibinfo{person}{Amjad Almahairi}, \bibinfo{person}{Yasmine Babaei},
  \bibinfo{person}{Nikolay Bashlykov}, \bibinfo{person}{Soumya Batra},
  \bibinfo{person}{Prajjwal Bhargava}, \bibinfo{person}{Shruti Bhosale},
  {et~al\mbox{.}}} \bibinfo{year}{2023}\natexlab{b}.
\newblock \showarticletitle{Llama 2: Open foundation and fine-tuned chat
  models}.
\newblock \bibinfo{journal}{{\em arXiv preprint arXiv:2307.09288\/}}
  (\bibinfo{year}{2023}).
\newblock


\bibitem[\protect\citeauthoryear{Touvron, Martin, Stone, Albert, Almahairi,
  Babaei, Bashlykov, Batra, Bhargava, Bhosale, Bikel, Blecher, Canton{-}Ferrer,
  Chen, Cucurull, Esiobu, Fernandes, Fu, Fu, Fuller, Gao, Goswami, Goyal,
  Hartshorn, Hosseini, Hou, Inan, Kardas, Kerkez, Khabsa, Kloumann, Korenev,
  Koura, Lachaux, Lavril, Lee, Liskovich, Lu, Mao, Martinet, Mihaylov, Mishra,
  Molybog, Nie, Poulton, Reizenstein, Rungta, Saladi, Schelten, Silva, Smith,
  Subramanian, Tan, Tang, Taylor, Williams, Kuan, Xu, Yan, Zarov, Zhang, Fan,
  Kambadur, Narang, Rodriguez, Stojnic, Edunov, and Scialom}{Touvron
  et~al\mbox{.}}{2023a}]%
        {DBLP:llama2}
\bibfield{author}{\bibinfo{person}{Hugo Touvron}, \bibinfo{person}{Louis
  Martin}, \bibinfo{person}{Kevin Stone}, \bibinfo{person}{Peter Albert},
  \bibinfo{person}{Amjad Almahairi}, \bibinfo{person}{Yasmine Babaei},
  \bibinfo{person}{Nikolay Bashlykov}, \bibinfo{person}{Soumya Batra},
  \bibinfo{person}{Prajjwal Bhargava}, \bibinfo{person}{Shruti Bhosale},
  \bibinfo{person}{Dan Bikel}, \bibinfo{person}{Lukas Blecher},
  \bibinfo{person}{Cristian Canton{-}Ferrer}, \bibinfo{person}{Moya Chen},
  \bibinfo{person}{Guillem Cucurull}, \bibinfo{person}{David Esiobu},
  \bibinfo{person}{Jude Fernandes}, \bibinfo{person}{Jeremy Fu},
  \bibinfo{person}{Wenyin Fu}, \bibinfo{person}{Brian Fuller},
  \bibinfo{person}{Cynthia Gao}, \bibinfo{person}{Vedanuj Goswami},
  \bibinfo{person}{Naman Goyal}, \bibinfo{person}{Anthony Hartshorn},
  \bibinfo{person}{Saghar Hosseini}, \bibinfo{person}{Rui Hou},
  \bibinfo{person}{Hakan Inan}, \bibinfo{person}{Marcin Kardas},
  \bibinfo{person}{Viktor Kerkez}, \bibinfo{person}{Madian Khabsa},
  \bibinfo{person}{Isabel Kloumann}, \bibinfo{person}{Artem Korenev},
  \bibinfo{person}{Punit~Singh Koura}, \bibinfo{person}{Marie{-}Anne Lachaux},
  \bibinfo{person}{Thibaut Lavril}, \bibinfo{person}{Jenya Lee},
  \bibinfo{person}{Diana Liskovich}, \bibinfo{person}{Yinghai Lu},
  \bibinfo{person}{Yuning Mao}, \bibinfo{person}{Xavier Martinet},
  \bibinfo{person}{Todor Mihaylov}, \bibinfo{person}{Pushkar Mishra},
  \bibinfo{person}{Igor Molybog}, \bibinfo{person}{Yixin Nie},
  \bibinfo{person}{Andrew Poulton}, \bibinfo{person}{Jeremy Reizenstein},
  \bibinfo{person}{Rashi Rungta}, \bibinfo{person}{Kalyan Saladi},
  \bibinfo{person}{Alan Schelten}, \bibinfo{person}{Ruan Silva},
  \bibinfo{person}{Eric~Michael Smith}, \bibinfo{person}{Ranjan Subramanian},
  \bibinfo{person}{Xiaoqing~Ellen Tan}, \bibinfo{person}{Binh Tang},
  \bibinfo{person}{Ross Taylor}, \bibinfo{person}{Adina Williams},
  \bibinfo{person}{Jian~Xiang Kuan}, \bibinfo{person}{Puxin Xu},
  \bibinfo{person}{Zheng Yan}, \bibinfo{person}{Iliyan Zarov},
  \bibinfo{person}{Yuchen Zhang}, \bibinfo{person}{Angela Fan},
  \bibinfo{person}{Melanie Kambadur}, \bibinfo{person}{Sharan Narang},
  \bibinfo{person}{Aur{\'{e}}lien Rodriguez}, \bibinfo{person}{Robert Stojnic},
  \bibinfo{person}{Sergey Edunov}, {and} \bibinfo{person}{Thomas Scialom}.}
  \bibinfo{year}{2023}\natexlab{a}.
\newblock \showarticletitle{Llama 2: Open Foundation and Fine-Tuned Chat
  Models}.
\newblock \bibinfo{journal}{{\em CoRR\/}}  \bibinfo{volume}{abs/2307.09288}
  (\bibinfo{year}{2023}).
\newblock


\bibitem[\protect\citeauthoryear{Vaswani, Shazeer, Parmar, Uszkoreit, Jones,
  Gomez, Kaiser, and Polosukhin}{Vaswani et~al\mbox{.}}{2017}]%
        {vaswani2017attention}
\bibfield{author}{\bibinfo{person}{Ashish Vaswani}, \bibinfo{person}{Noam
  Shazeer}, \bibinfo{person}{Niki Parmar}, \bibinfo{person}{Jakob Uszkoreit},
  \bibinfo{person}{Llion Jones}, \bibinfo{person}{Aidan~N Gomez},
  \bibinfo{person}{{\L}ukasz Kaiser}, {and} \bibinfo{person}{Illia
  Polosukhin}.} \bibinfo{year}{2017}\natexlab{}.
\newblock \showarticletitle{Attention is all you need}.
\newblock \bibinfo{journal}{{\em Neural Information Processing Systems\/}}
  (\bibinfo{year}{2017}).
\newblock


\bibitem[\protect\citeauthoryear{Verma}{Verma}{2021}]%
        {DBLP:journals/corr/abs-2101-10277}
\bibfield{author}{\bibinfo{person}{Madhusudan Verma}.}
  \bibinfo{year}{2021}\natexlab{}.
\newblock \showarticletitle{Revisiting Linformer with a modified self-attention
  with linear complexity}.
\newblock \bibinfo{journal}{{\em CoRR\/}}  \bibinfo{volume}{abs/2101.10277}
  (\bibinfo{year}{2021}).
\newblock


\bibitem[\protect\citeauthoryear{Wu, Liu, Zhong, Sun, Liu, and Jin}{Wu
  et~al\mbox{.}}{2024}]%
        {DBLP:journals/corr/abs-2404-09526}
\bibfield{author}{\bibinfo{person}{Bingyang Wu}, \bibinfo{person}{Shengyu Liu},
  \bibinfo{person}{Yinmin Zhong}, \bibinfo{person}{Peng Sun},
  \bibinfo{person}{Xuanzhe Liu}, {and} \bibinfo{person}{Xin Jin}.}
  \bibinfo{year}{2024}\natexlab{}.
\newblock \showarticletitle{LoongServe: Efficiently Serving Long-context Large
  Language Models with Elastic Sequence Parallelism}.
\newblock \bibinfo{journal}{{\em CoRR\/}}  \bibinfo{volume}{abs/2404.09526}
  (\bibinfo{year}{2024}).
\newblock


\bibitem[\protect\citeauthoryear{Wu, Zhong, Zhang, Huang, Liu, and Jin}{Wu
  et~al\mbox{.}}{2023a}]%
        {DBLP:journals/corr/abs-2305-05920}
\bibfield{author}{\bibinfo{person}{Bingyang Wu}, \bibinfo{person}{Yinmin
  Zhong}, \bibinfo{person}{Zili Zhang}, \bibinfo{person}{Gang Huang},
  \bibinfo{person}{Xuanzhe Liu}, {and} \bibinfo{person}{Xin Jin}.}
  \bibinfo{year}{2023}\natexlab{a}.
\newblock \showarticletitle{Fast Distributed Inference Serving for Large
  Language Models}.
\newblock \bibinfo{journal}{{\em CoRR\/}}  \bibinfo{volume}{abs/2305.05920}
  (\bibinfo{year}{2023}).
\newblock


\bibitem[\protect\citeauthoryear{Wu, Zhong, Zhang, Huang, Liu, and Jin}{Wu
  et~al\mbox{.}}{2023b}]%
        {wu2023fast}
\bibfield{author}{\bibinfo{person}{Bingyang Wu}, \bibinfo{person}{Yinmin
  Zhong}, \bibinfo{person}{Zili Zhang}, \bibinfo{person}{Gang Huang},
  \bibinfo{person}{Xuanzhe Liu}, {and} \bibinfo{person}{Xin Jin}.}
  \bibinfo{year}{2023}\natexlab{b}.
\newblock \showarticletitle{Fast Distributed Inference Serving for Large
  Language Models}.
\newblock \bibinfo{journal}{{\em arXiv\/}} (\bibinfo{year}{2023}).
\newblock


\bibitem[\protect\citeauthoryear{Xiao, Tian, Chen, Han, and Lewis}{Xiao
  et~al\mbox{.}}{2024}]%
        {DBLP:streaming-llm}
\bibfield{author}{\bibinfo{person}{Guangxuan Xiao}, \bibinfo{person}{Yuandong
  Tian}, \bibinfo{person}{Beidi Chen}, \bibinfo{person}{Song Han}, {and}
  \bibinfo{person}{Mike Lewis}.} \bibinfo{year}{2024}\natexlab{}.
\newblock \showarticletitle{Efficient Streaming Language Models with Attention
  Sinks}. In \bibinfo{booktitle}{{\em The Twelfth International Conference on
  Learning Representations, {ICLR} 2024, Vienna, Austria, May 7-11, 2024}}.
  \bibinfo{publisher}{OpenReview.net}.
\newblock


\bibitem[\protect\citeauthoryear{Xu, Zhang, Guo, Hu, Liu, Wu, Feng, Sun, Shao,
  Guo, Zhao, Zhang, Guo, and Leng}{Xu et~al\mbox{.}}{2024}]%
        {xu2024vtensor}
\bibfield{author}{\bibinfo{person}{Jiale Xu}, \bibinfo{person}{Rui Zhang},
  \bibinfo{person}{Cong Guo}, \bibinfo{person}{Weiming Hu},
  \bibinfo{person}{Zihan Liu}, \bibinfo{person}{Feiyang Wu},
  \bibinfo{person}{Yu Feng}, \bibinfo{person}{Shixuan Sun},
  \bibinfo{person}{Changxu Shao}, \bibinfo{person}{Yuhong Guo},
  \bibinfo{person}{Junping Zhao}, \bibinfo{person}{Ke Zhang},
  \bibinfo{person}{Minyi Guo}, {and} \bibinfo{person}{Jingwen Leng}.}
  \bibinfo{year}{2024}\natexlab{}.
\newblock \showarticletitle{vTensor: Flexible Virtual Tensor Management for
  Efficient LLM Serving}.
\newblock \bibinfo{journal}{{\em arXiv\/}} (\bibinfo{year}{2024}).
\newblock


\bibitem[\protect\citeauthoryear{Young, Chen, Li, Huang, Zhang, Zhang, Li, Zhu,
  Chen, Chang, Yu, Liu, Liu, Yue, Yang, Yang, Yu, Xie, Huang, Hu, Ren, Niu,
  Nie, Xu, Liu, Wang, Cai, Gu, Liu, and Dai}{Young et~al\mbox{.}}{2024}]%
        {DBLP:Yi}
\bibfield{author}{\bibinfo{person}{Alex Young}, \bibinfo{person}{Bei Chen},
  \bibinfo{person}{Chao Li}, \bibinfo{person}{Chengen Huang},
  \bibinfo{person}{Ge Zhang}, \bibinfo{person}{Guanwei Zhang},
  \bibinfo{person}{Heng Li}, \bibinfo{person}{Jiangcheng Zhu},
  \bibinfo{person}{Jianqun Chen}, \bibinfo{person}{Jing Chang},
  \bibinfo{person}{Kaidong Yu}, \bibinfo{person}{Peng Liu},
  \bibinfo{person}{Qiang Liu}, \bibinfo{person}{Shawn Yue},
  \bibinfo{person}{Senbin Yang}, \bibinfo{person}{Shiming Yang},
  \bibinfo{person}{Tao Yu}, \bibinfo{person}{Wen Xie}, \bibinfo{person}{Wenhao
  Huang}, \bibinfo{person}{Xiaohui Hu}, \bibinfo{person}{Xiaoyi Ren},
  \bibinfo{person}{Xinyao Niu}, \bibinfo{person}{Pengcheng Nie},
  \bibinfo{person}{Yuchi Xu}, \bibinfo{person}{Yudong Liu},
  \bibinfo{person}{Yue Wang}, \bibinfo{person}{Yuxuan Cai},
  \bibinfo{person}{Zhenyu Gu}, \bibinfo{person}{Zhiyuan Liu}, {and}
  \bibinfo{person}{Zonghong Dai}.} \bibinfo{year}{2024}\natexlab{}.
\newblock \showarticletitle{Yi: Open Foundation Models by 01.AI}.
\newblock \bibinfo{journal}{{\em CoRR\/}}  \bibinfo{volume}{abs/2403.04652}
  (\bibinfo{year}{2024}).
\newblock


\bibitem[\protect\citeauthoryear{Yu, Jeong, Kim, Kim, and Chun}{Yu
  et~al\mbox{.}}{2022}]%
        {DBLP:orca}
\bibfield{author}{\bibinfo{person}{Gyeong{-}In Yu}, \bibinfo{person}{Joo~Seong
  Jeong}, \bibinfo{person}{Geon{-}Woo Kim}, \bibinfo{person}{Soojeong Kim},
  {and} \bibinfo{person}{Byung{-}Gon Chun}.} \bibinfo{year}{2022}\natexlab{}.
\newblock \showarticletitle{Orca: {A} Distributed Serving System for
  Transformer-Based Generative Models}. In \bibinfo{booktitle}{{\em 16th
  {USENIX} Symposium on Operating Systems Design and Implementation, {OSDI}
  2022, Carlsbad, CA, USA, July 11-13, 2022}},
  \bibfield{editor}{\bibinfo{person}{Marcos~K. Aguilera} {and}
  \bibinfo{person}{Hakim Weatherspoon}} (Eds.). \bibinfo{publisher}{{USENIX}
  Association}, \bibinfo{pages}{521--538}.
\newblock


\bibitem[\protect\citeauthoryear{Zhang, Sheng, Zhou, Chen, Zheng, Cai, Song,
  Tian, R{\'e}, Barrett, et~al\mbox{.}}{Zhang et~al\mbox{.}}{2024}]%
        {zhang2023h2o}
\bibfield{author}{\bibinfo{person}{Zhenyu Zhang}, \bibinfo{person}{Ying Sheng},
  \bibinfo{person}{Tianyi Zhou}, \bibinfo{person}{Tianlong Chen},
  \bibinfo{person}{Lianmin Zheng}, \bibinfo{person}{Ruisi Cai},
  \bibinfo{person}{Zhao Song}, \bibinfo{person}{Yuandong Tian},
  \bibinfo{person}{Christopher R{\'e}}, \bibinfo{person}{Clark Barrett},
  {et~al\mbox{.}}} \bibinfo{year}{2024}\natexlab{}.
\newblock \showarticletitle{H2o: Heavy-hitter oracle for efficient generative
  inference of large language models}. In \bibinfo{booktitle}{{\em Neural
  Information Processing Systems}}.
\newblock


\bibitem[\protect\citeauthoryear{Zhang, Sheng, Zhou, Chen, Zheng, Cai, Song,
  Tian, R{\'{e}}, Barrett, Wang, and Chen}{Zhang et~al\mbox{.}}{2023}]%
        {DBLP:h2o}
\bibfield{author}{\bibinfo{person}{Zhenyu Zhang}, \bibinfo{person}{Ying Sheng},
  \bibinfo{person}{Tianyi Zhou}, \bibinfo{person}{Tianlong Chen},
  \bibinfo{person}{Lianmin Zheng}, \bibinfo{person}{Ruisi Cai},
  \bibinfo{person}{Zhao Song}, \bibinfo{person}{Yuandong Tian},
  \bibinfo{person}{Christopher R{\'{e}}}, \bibinfo{person}{Clark~W. Barrett},
  \bibinfo{person}{Zhangyang Wang}, {and} \bibinfo{person}{Beidi Chen}.}
  \bibinfo{year}{2023}\natexlab{}.
\newblock \showarticletitle{{H2O:} Heavy-Hitter Oracle for Efficient Generative
  Inference of Large Language Models}. In \bibinfo{booktitle}{{\em Advances in
  Neural Information Processing Systems 36: Annual Conference on Neural
  Information Processing Systems 2023, NeurIPS 2023, New Orleans, LA, USA,
  December 10 - 16, 2023}}, \bibfield{editor}{\bibinfo{person}{Alice Oh},
  \bibinfo{person}{Tristan Naumann}, \bibinfo{person}{Amir Globerson},
  \bibinfo{person}{Kate Saenko}, \bibinfo{person}{Moritz Hardt}, {and}
  \bibinfo{person}{Sergey Levine}} (Eds.).
\newblock


\bibitem[\protect\citeauthoryear{Zheng, Li, Zhang, Zhuang, Chen, Huang, Wang,
  Xu, Zhuo, Xing, Gonzalez, and Stoica}{Zheng et~al\mbox{.}}{2022}]%
        {DBLP:alpa}
\bibfield{author}{\bibinfo{person}{Lianmin Zheng}, \bibinfo{person}{Zhuohan
  Li}, \bibinfo{person}{Hao Zhang}, \bibinfo{person}{Yonghao Zhuang},
  \bibinfo{person}{Zhifeng Chen}, \bibinfo{person}{Yanping Huang},
  \bibinfo{person}{Yida Wang}, \bibinfo{person}{Yuanzhong Xu},
  \bibinfo{person}{Danyang Zhuo}, \bibinfo{person}{Eric~P. Xing},
  \bibinfo{person}{Joseph~E. Gonzalez}, {and} \bibinfo{person}{Ion Stoica}.}
  \bibinfo{year}{2022}\natexlab{}.
\newblock \showarticletitle{Alpa: Automating Inter- and Intra-Operator
  Parallelism for Distributed Deep Learning}. In \bibinfo{booktitle}{{\em 16th
  {USENIX} Symposium on Operating Systems Design and Implementation, {OSDI}
  2022, Carlsbad, CA, USA, July 11-13, 2022}},
  \bibfield{editor}{\bibinfo{person}{Marcos~K. Aguilera} {and}
  \bibinfo{person}{Hakim Weatherspoon}} (Eds.). \bibinfo{publisher}{{USENIX}
  Association}, \bibinfo{pages}{559--578}.
\newblock


\bibitem[\protect\citeauthoryear{Zhong, Liu, Chen, Hu, Zhu, Liu, Jin, and
  Zhang}{Zhong et~al\mbox{.}}{2024}]%
        {zhong2024distserve}
\bibfield{author}{\bibinfo{person}{Yinmin Zhong}, \bibinfo{person}{Shengyu
  Liu}, \bibinfo{person}{Junda Chen}, \bibinfo{person}{Jianbo Hu},
  \bibinfo{person}{Yibo Zhu}, \bibinfo{person}{Xuanzhe Liu},
  \bibinfo{person}{Xin Jin}, {and} \bibinfo{person}{Hao Zhang}.}
  \bibinfo{year}{2024}\natexlab{}.
\newblock \showarticletitle{DistServe: Disaggregating Prefill and Decoding for
  Goodput-optimized Large Language Model Serving}.
\newblock \bibinfo{journal}{{\em arXiv\/}} (\bibinfo{year}{2024}).
\newblock


\end{thebibliography}

\clearpage


\end{document}